\documentclass[aps,twocolumn,prb,floatfix,reprint,twocolumn,superscriptaddress,amsmath,amssymb]{revtex4-1}
\pdfoutput=1
\usepackage[latin9]{inputenc}
\usepackage{amsmath}
\usepackage{graphicx}
\usepackage[bookmarks=false,linkcolor=blue,urlcolor=blue,colorlinks,citecolor=blue]{hyperref}


\graphicspath{{./}} % for pdf_svg

\begin{document}

\global\long\def\gvec#1{\boldsymbol{#1}}
\global\long\def\bra#1{\langle#1|}
\global\long\def\ket#1{|#1\rangle}
\global\long\def\braket#1#2{\langle#1|#2\rangle}
\global\long\def\avg#1{\langle#1\rangle}

\title{Universal phase diagram of topological superconductors subjected to magnetic flux}

\author{Omri Lesser}
\author{Yuval Oreg}

\affiliation{Department of Condensed Matter Physics, Weizmann Institute of Science,
Rehovot 7610001, Israel}
\begin{abstract}
We perform a theoretical study of the orbital effect of a magnetic field on a proximity-coupled islands array of $p_{x}+ip_{y}$ topological superconductors.
To describe the system, we generalize the tight-binding model of the Hofstadter butterfly to include the effect of the superconducting islands.
The quantum Hall topological phases, appearing in  the absence of superconductivity, are characterized by integer fermionic Chern numbers corresponding to the number of occupied bulk Landau levels.
As the strength of the superconducting pairing increases a series of transitions occurs, with one less chiral Majorana edge mode at each consecutive phase, leading to a reduction of the fermionic Chern number by a half.
When the pairing potential exceeds the tight-binding model bandwidth, Cooper pairs are localized in the islands, the Chern number is zero, and there are no low-energy edge modes.
We identify domains in the model's parameter space for which the system is topological and supports an odd number of chiral Majorana edge modes.
While the precise shape of the domains depends on the details of the model, the general structure of the phase diagram is robust, and it is obtained numerically and in several simplified traceable analytical models.
We discuss the relevance of this study to recent experimental research of two-dimensional superconductor arrays on semiconductor systems.
\end{abstract}

\date{\today}


\maketitle

\section{Introduction}

Low-dimensional topological superconductors (SCs) have been the subject of vast interest recently~\cite{alicea_new_2012,bernevig_topological_2013}.
That is because they constitue a novel form of quantum matter, and also because they may support Majorana zero modes, which are germane for non-Abelian topological quantum processing~\cite{nayak_non-abelian_2008}.

A time-reversal symmetry breaking topological superconductor is realized when an odd number of energy bands, which are not spin degenerate, are unstable to the creation of Cooper pairs, while the other bands are gapped.

This may be obtained by the proper combination of materials that have strong spin-orbit coupling, are proximity coupled to a superconductor, and break time-reversal symmetry by internal magnetic phenomena~\cite{sau_generic_2010,palacio-morales_atomic-scale_2019} or external application of a magnetic field~\cite{lutchyn_majorana_2010,oreg_helical_2010}.
Several realization schemes for topological superconductivity have been put forward in two spatial dimensions~\cite{albuquerque_engineering_2008,alicea_majorana_2010,fu_superconducting_2008,sau_generic_2010,qi_chiral_2010}.

Application of an external magnetic field affects electrons both through the orbital degrees of freedom and through the Zeeman term  coupled to the spin.
For systems with a large $g$ factor the latter is more significant, especially when the magnetic field is parallel to the wire's axis in the one-dimensional case \footnote{Full shell systems are an exception; see Ref.~\cite{lutchyn_topological_2018}} or to the system's plane in the two-dimensional case.
However, when the magnetic field is applied perpendicular to a two-dimensional system, we expect an interesting interplay between the (quantum) Hall effect~\cite{klitzing_new_1980,stern_anyons_2008,girvin_quantum_1999} and the emergence of topological superconductivity, decorated by vortices induced by the perpendicular field.

\begin{figure}
\begin{centering}
\def\svgwidth{0.9\linewidth}
{\small 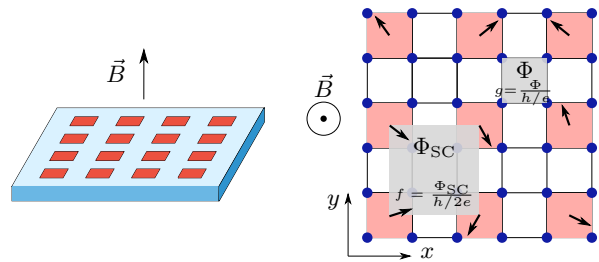}
\par\end{centering}
\caption{\label{fig:islands_system}
Illustration of the islands array: a 3D view is shown in the left panel and a top view (of a part of the system) is depicted in the right panel.
Superconductivity exists only in the red squares.
In the right panel blue circles denote lattice sites of a tight-binding model of the system.
Nearest-neighbor hopping is allowed and uniform throughout the entire lattice.
Inside each island, the phase of the superconducting order parameter is approximately constant and its direction is illustrated by an arrow.
The magnetic flux per plaquette is $\Phi$, and the flux per SC plaquette is $\Phi_{\text{SC}}$, as illustrated by the gray squares.
We define the parameters $g=\frac{\Phi}{h/e}$ which is the flux per unit plaquette and $f=\frac{\Phi_{\text{SC}}}{h/2e}$ which is the flux per SC plaquette. In the geometry depicted here, $f=8g$ (see Appendix~\ref{appendix:geometry}).}
\end{figure}

To access this interplay, we study a tight-binding model which gives rise to Hofstadter-butterfly physics~\cite{harper_single_1955,hofstadter_energy_1976}, with proximity coupling to an array of SC islands (see Fig.~\ref{fig:islands_system}).
Whereas the parts that are not covered by the SC are expected to be susceptible to an external gate potential, sites that are in proximity to the SC islands are less sensitive to the gate potential but are coupled via a pairing potential.
This setup, inspired by a recent experiment~\cite{bottcher_superconducting_2018}, enables proximity effects together with tunability of the chemical potential.

The phase of the SC order parameter is assumed to be constant inside each island.
This is a good approximation as long as the flux per plaquette $\Phi$ is much smaller than the flux quantum $\Phi_{0}=h/e$, since in this regime the SC phase does not wind inside an island.
To establish the regime of validity of this approximation, we note that the flux piercing an island of size $L\times L$ when applying a perpendicular magnetic field $B$ is  $\Phi/\Phi_0 = B\text{[T]} \cdot \left(L\text{[nm]}\right)^2 \cdot 2.4\times 10^{-4}$.
Typically a magnetic field of the order of $1\text{T}$ is required to get a substantial Zeeman splitting and tune into the topological SC phase, so $L$ must be smaller than $\sim 64\text{nm}$.
We note that in such small sizes the Coulomb energy of an isolated island may be large; however, if the island is thick and the effective coupling between the islands is strong, this effect may be neglected~\cite{aleiner_quantum_2002}, as is done throughout this paper.

The system we investigate may be viewed as the superconducting generalization of Hofstadter's model.
In addition to the known commensurability effects which are related to the magnetic unit cell, here the geometry of the superlattice plays a role.
Namely, in addition to the unit plaquette there is also a SC unit plaquette (see Fig.~\ref{fig:islands_system}).
The orbital effect of the magnetic field modifies the state of the SC, and therefore the ground-state phase configuration of the SC order parameter is not uniform.
We find this phase-configuration numerically and also by drawing an analogy to the simpler, exactly-solvable frustrated XY model~\cite{jos_40_2013}.

Our model describes a $p_{x}+ip_{y}$ SC, with an application of a non-zero perpendicular magnetic field.
Besides being interesting on its own merit, this model has experimental relevance as the orbital effects are almost inevitable \footnote{they can be eliminated if the magnetic field is applied in-plane and strong Dresselhaus spin-orbit coupling (like in (111) InSb samples~\cite{alicea_majorana_2010}) is present~\cite{dresselhaus_spin-orbit_1955,levine_realizing_2017}}.
The magnetic field may induce vortices in the SC (or between the SC islands); since the $p_{x}+ip_{y}$ SC can be tuned into a topologically non-trivial phase (i.e. one that supports chiral Majorana edge modes), a Majorana bound state is expected to reside at the core of each vortex.
In the effective low-energy description of the system, the matrix elements between the Majorana bound states at close vortices are non-zero.
The coupling between them may give rise to a chiral Majorana mode at the edge of the whole sample.

In this manuscript, we show that Majorana edge modes indeed appear, and the system can be driven into a global topological phase, by the application of an external perpendicular field.
This is formally shown in Sec.~\ref{Sec:SysAnalysis} by calculating the topological invariant of symmetry class D in two dimensions -- the Chern number\cite{altland_nonstandard_1997,
schnyder_classification_2008,
kitaev_periodic_2009} $\mathcal{N}$, which is by definition equal to the number of chiral Majorana edge modes.
When this Chern number is even, two Majorana modes can pair up to form one fermion edge mode, and the Chern number $\mathcal{N}/2$ is similar to that of the integer quantum Hall effect.
However when the Chern number is odd, one unpaired Majorana edge mode exists, and the system enters the topological superconducting state.

Our main result is the phase diagram shown in Fig.~\ref{fig:chern_phase_diagram}, where the parameter space includes substantial regions with odd Chern number.
This diagram exhibits a series of quantum Hall transitions, where the Chern number ${\cal N}$ jumps in steps of two, as we sweep  the chemical potential keeping the pairing potential $\Delta=0$. At finite $\Delta$  the jumps split to two ``Majorana transitions" in which the Chern number changes by one.
At large $\Delta$ the system becomes topologically trivial, which we attribute to the localization of Cooper pairs in islands.

We then turn to examine the universal features of the phase diagram in Sec.~\ref{Sec:GenericForm}.
We show that even though the system under study is very rich and complex, and depends on many parameters, its phase diagram shares many generic features with the phase diagrams of much simpler models.
The fact that the SC pairing potential is staggered is found in Subsec.~\ref{subsec:effect-of-staggering} to be important to yield zero Chern number for large pairing potential.
The universal properties are understood by considering the symmetries of such models, which depend (for a fixed external field) on two parameters, e.g.,  $p_1$ and $p_2$, the analogs of $\Delta$ and $\mu$.
When $p_1=0$, a sweep of $p_2$ induces transitions between even Chern numbers in steps of two.
Relieving the constraint on $p_1$ splits the transitions, and exposes regions with odd Chern numbers.
As $p_1$ increases the Chern number decreases in unit steps until it vanishes for large $p_1$, making the system topologically trivial.
To elucidate these features, we study a simplified version of the 2D model in Subsec.~\ref{subsec:Semi-analytical-2D}, and a generalization of Kitaev's chain model~\cite{kitaev_unpaired_2001} in Subsec.~\ref{subsec:Stacking-Kitaev-chains}.
These two models yield phase diagrams which are similar to that of the full 2D model (see Figs.~\ref{fig:stripes_phase_diagram}, \ref{fig:coupled_Kitaev_phase_diagram}).

We conclude with a discussion of the main results of the paper.
We comment on possible realizations of the effective $p$-wave pairing we have assumed here in systems with spin-orbit coupling, and ways to detect the topological phase using heat transport.

Details of the calculations are relegated to the following appendices.
In Appendix~\ref{appendix:ground_state} we provide additional details on the numerical approach used to find the ground-state phase configuration.
In Appendix~\ref{appendix:geometry} we elaborate on the geometrical properties of the superlattice.
In Appendix~\ref{appendix:chern_realspace} we review the method used to calculate the Chern number in real space.
In Appendix~\ref{appendix:additional_2d_results} we present several additional numerical results in the 2D system to support our main results.

\section{Analysis of the system}\label{Sec:SysAnalysis}

\subsection{Model}

We introduce the following tight-binding Hamiltonian for single-species fermions on a two-dimensional square lattice, with superimposed $p_x+ip_y$ SC islands:
\begin{equation}
\begin{aligned}H= & -\mu\sum_{m,n}c_{m,n}^{\dagger}c_{m,n} + \\
 & \left[-t\sum_{m,n}\left(c_{m,n}^{\dagger}c_{m+1,n}+e^{-2\pi im\frac{\Phi}{\Phi_{0}}}c_{m,n}^{\dagger}c_{m,n+1}\right) \right.\\
 & +\Delta\sum_{j\in\text{islands}}e^{i\theta_{j}}\sum_{m,n\in j^{\text{th}}\text{ island}}\left(c_{m,n}c_{m+1,n}\right.\\
 & \left.+ic_{m,n}c_{m,n+1}\right) + \text{h.c.} \Bigg].
\end{aligned}
\label{eq:islands_hamiltonian}
\end{equation}
Here
$c_{m,n}^{\dagger},c_{m,n}$ are creation and annihilation operators of fermions in the lattice site labelled by $(m,n)$, whose location is $a\left(m\hat{x}+n\hat{y}\right)$ where $a$ is the lattice constant;
$\mu$ is the chemical potential;
$t$ is the hopping amplitude between nearest-neighboring sites; % which is related to a 2DEG model by t=\frac{1}{2m_{e}a^{2}}
$\Phi$ is the flux per unit plaquette, which enters the Hamiltonian via the Peierls substitution~\cite{peierls_zur_1933} using the Landau gauge $\vec{A}=\Phi x\hat{y}/a^2$;
$\Delta$ is the magnitude of the induced superconducting pairing potential;
and $\theta_{j}$ is the phase of the superconducting order parameter in the $j^{\text{th}}$ island, which is approximately constant inside the island.

The Hamiltonian can be brought to the Bogoliuobv-de-Gennes (BdG) form by defining the Nambu spinor
\begin{equation}
\Psi = \left( c_{1,1} , c_{1,2} , \ldots , c_{2,1} , \ldots c_{N_x,N_y} , c^{\dagger}_{1,1} , \ldots , c^{\dagger}_{N_x,N_y} \right)^T
\end{equation}
and writing
\begin{equation}
\begin{gathered}
    H = \Psi^{\dagger} \mathcal{H}_{\mathrm{BdG}} \Psi, \\
    \mathcal{H}_{\mathrm{BdG}} = \begin{pmatrix} H_0 && H_{\Delta} \\ H_{\Delta}^{\dagger} && -H_0^T \end{pmatrix},
\end{gathered}
\label{eq:H_BdG}
\end{equation}
where $H_0$ corresponds to the normal $\mu,t$ terms and $H_{\Delta}=-H_{\Delta}^T$ corresponds to the SC terms.
$\mathcal{H}_{\mathrm{BdG}}$ has particle-hole symmetry $\left\{ \mathcal{H}_{\mathrm{BdG}},\Lambda\right\} =0$ with $\Lambda=\tau_{x}\mathcal{K}$, where $\tau_x$ is a Pauli matrix acting in particle-hole space and $\mathcal{K}$ is complex conjugation.
The magnetic field and the chiral $p$-wave pairing terms break time-reversal symmetry $\left[\mathcal{H}_{\mathrm{BdG}},\mathcal{K}\right]\neq 0$, placing $\mathcal{H}_{\mathrm{BdG}}$ in symmetry class D with a $\mathbb{Z}$ topological invariant\cite{altland_nonstandard_1997,
schnyder_classification_2008,
kitaev_periodic_2009}.

In the Hamiltonian Eq.~\eqref{eq:islands_hamiltonian}, the phases $\{\theta_{j}\}$ are treated as parameters.
In reality, due to the interaction between the SC and the magnetic field, the ground state configuration of these phases should be determined in a self-consistent way.
This can be done numerically, as elaborated in Appendix~\ref{appendix:ground_state}.
Due to the computational complexity of this method, we employed a simpler, approximate way to obtain the ground state phase configuration.
We drawing an analogy between the current system and the two-dimensional frustrated XY model~\cite{jos_40_2013} -- a classical theory of $U(1)$ spins with a nearest-neighbor exchange interaction in the presence of a gauge field.
Our model is composed of an array of Josephson junctions, and therefore, upon integrating out the fermionic degrees of freedom, the low-energy effective theory for the bosonic SC phases $\{\theta_{j}\}$ has the same symmetry properties as the frustrated XY model.
The effective flux per SC plaquette $f=\frac{\Phi_{\text{SC}}}{h/2e}$ is  related to the actual flux per plaquette $g=\frac{\Phi}{h/e}$ by geometric factors (see Appendix~\ref{appendix:geometry} for elaboration); concretely, $f=8g$ for the geometry of Fig.~\ref{fig:islands_system}.
We therefore substitute the known ground state of the frustrated XY model~\cite{halsey_josephson-junction_1985} with flux $f$ as the phase configuration of our system.

\subsection{Chern number}\label{subsec:Chern}
Having established the ground state configuration, we shall henceforth treat the phases $\{\theta_{j}\}$ in Eq.~\eqref{eq:islands_hamiltonian} as fixed parameters and study the properties of the resulting Hamiltonian.
A direct way to classify the topological properties of the system is calculating the $\mathbb{Z}$ topological invariant -- the Chern number~\cite{altland_nonstandard_1997,
schnyder_classification_2008,
kitaev_periodic_2009}.
Upon casting the Hamiltonian in BdG form, the Chern number $\mathcal{N}$ counts the number of chiral Majorana modes at the edge of the sample; an odd Chern number therefore corresponds to an unpaired chiral Majorana edge mode, which is the hallmark of topological superconductivity.

\begin{figure}%[ht]
\includegraphics[width=0.95\linewidth]{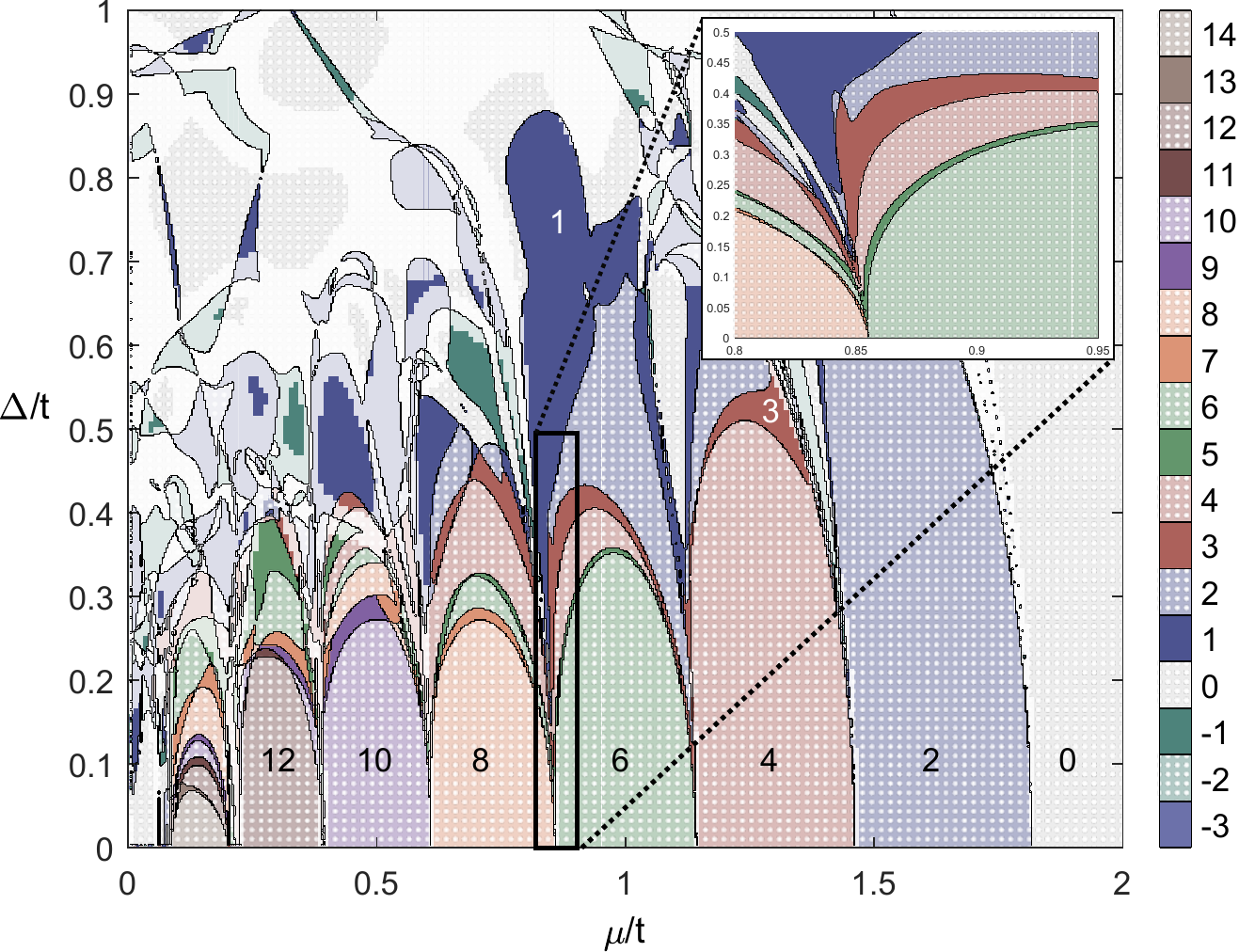}
\caption{\label{fig:chern_phase_diagram}
Chern number phase diagram for $f=\frac{1}{2}$ flux quantum per superconducting plaquette, as a function of the chemical potential $\mu$ and the induced SC pairing potential $\Delta$.
The colors indicate the Chern number: dark colors represent odd Chern numbers whereas light colors (with white dots) represent even Chern numbers.
At small $\Delta$ and at large $\Delta$ the system is in a trivial phase (having an even Chern number), whereas for some regions of intermediate $\Delta$ the system can be driven into a topological SC phase (having an odd Chern number) by tuning $\mu$.
Finite $\Delta$ splits the normal transitions, which are two-fold jumps of the Chern number, thereby creating slivers of odd Chern number.
Regions of small energy gap (smaller than the level spacing given by the bandwidth $t$ divided by the total number of sites) are shaded.
Inset: focus on one of the transitions, showing the splitting of the $8\rightarrow 6$ transition by finite $\Delta$, such that all the intermediate Chern numbers between 0 and 8 appear.
}
\end{figure}

The Chern number may be calculated in momentum space~\cite{fukui_chern_2005} or in real space~\cite{yi-fu_coupling-matrix_2013,loring_disordered_2011}.
In this work we used a real-space calculation, since it is computationally more efficient and numerically robust (the details of the real-space calculation are explained in Appendix~\ref{appendix:chern_realspace}).
In particular, since the SC phases vary in space, going to momentum space is not very beneficial.
All numerical calculation schemes are prone to finite size or resolution effects;
we minimized these by using large enough systems such that in the absence of superconductivity (i.e. $\Delta=0$) we obtained the known Chern numbers for the original Hofstadter model~\cite{thouless_quantized_1982}.
We then scanned the parameter space by varying $\mu$ and $\Delta$ ($t$ is held constant, fixing the energy scale), and calculated the Chern number for each set of parameters.
The resulting phase diagram for $f=\frac{1}{2}$ is shown in Fig.~\ref{fig:chern_phase_diagram}.

Several important features appear in the phase diagram Fig.~\ref{fig:chern_phase_diagram}.
At $\Delta=0$ we recover the original Hofstadter model, which gives a series of two-fold transitions of the Chern number as a function of $\mu$.
Recall that in the BdG form, the Chern number counts the number of Majorana edge modes, so a Chern number $C$ in the Hofstadter model becomes a BdG Chern number $\mathcal{N}=2C$.
The region of interest is intermediate $\Delta$, where a topological phase (odd Chern number) is stabilized in large areas of the parameter space, making it robust to small fluctuations in the parameters.

At large $\Delta$ the Chern number tends to zero.
The reason for this is the islands structure, namely that the SC covering is partial.
For large $\Delta$, such a structure gives rise to localized pairs which only realize a topologically trivial phase.
This argument is further supported in Subsec.~\ref{subsec:effect-of-staggering}, where we study the effect of SC staggering analytically.

In order to complement the Chern number phase diagram we examined the energy gap as a function of $\mu$ and $\Delta$ (see Fig.~\ref{fig:energy_gap_map} in Appendix~\ref{appendix:additional_2d_results}), and found that near $\mu=0$ and at finite $\Delta$ the system becomes gapless.
This is because for $\mu=0$ and $\Delta=0$ the system undergoes a transition of a large Chern number difference, resulting in a very small gap.
Finite $\Delta$ then mixes the states near $\mu=0$, resulting in a gapless state.
Notice that the Chern number is ill-defined in such a gapless state, so the area of the phase diagram close to $\mu=0$ is unreliable.
The method we used to calculate the Chern number always yields and integer, as the system is finite so a gap always exists; however, this integer number has no real physical meaning.
Therefore, in Fig.~\ref{fig:chern_phase_diagram} we shade regions where the energy gap is smaller than the finite-size level spacing, estimated by the tight-binding bandwidth $t$ divided by the total number of sites.
For completeness we show the Chern numbers of these regions in the phase diagram, although the true behavior there is metallic.
\begin{figure}[t]
\includegraphics[width=0.9\linewidth]{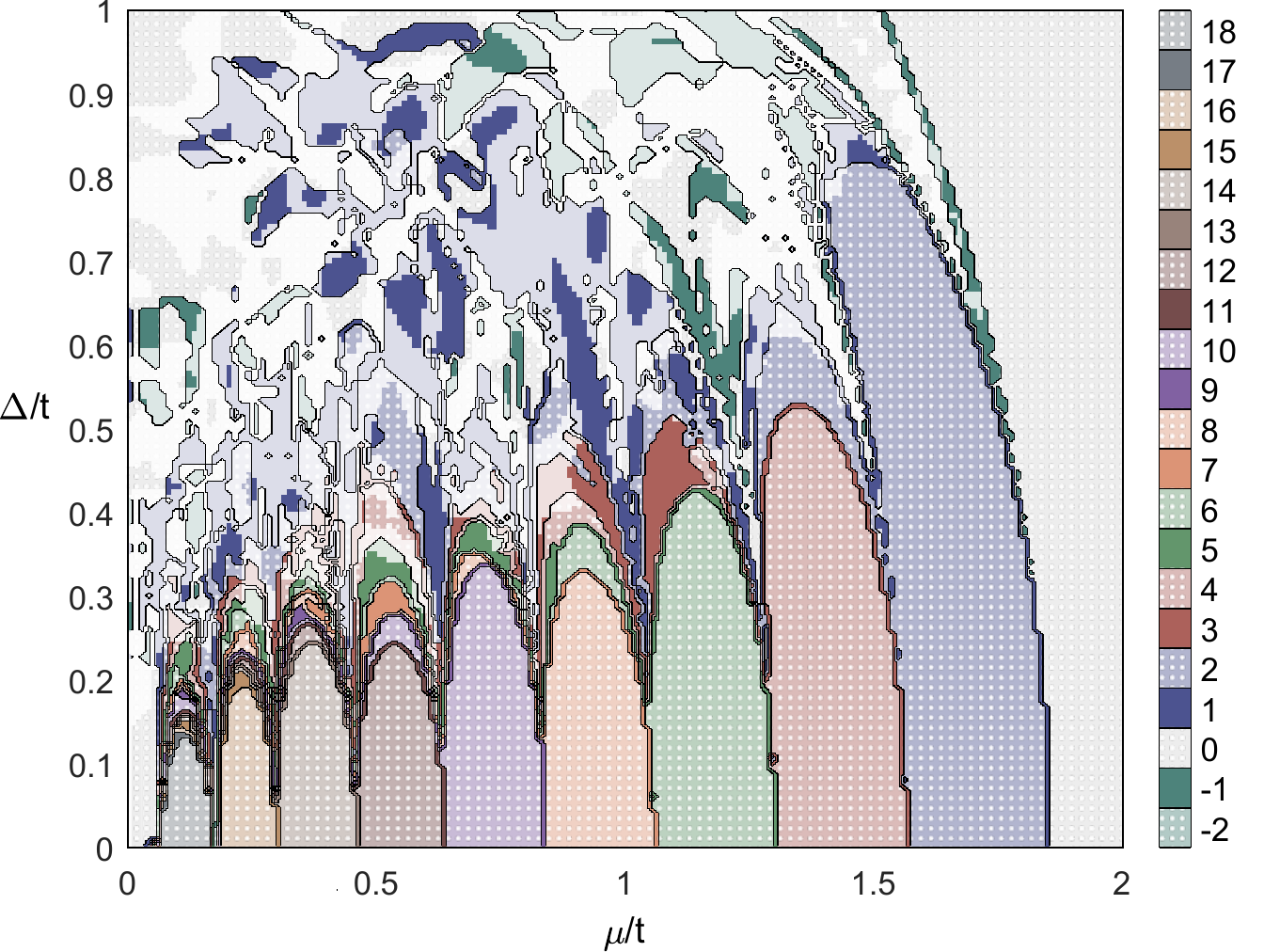}\\
(a) $f=2/5$\\$ $\\
\includegraphics[width=0.9\linewidth]{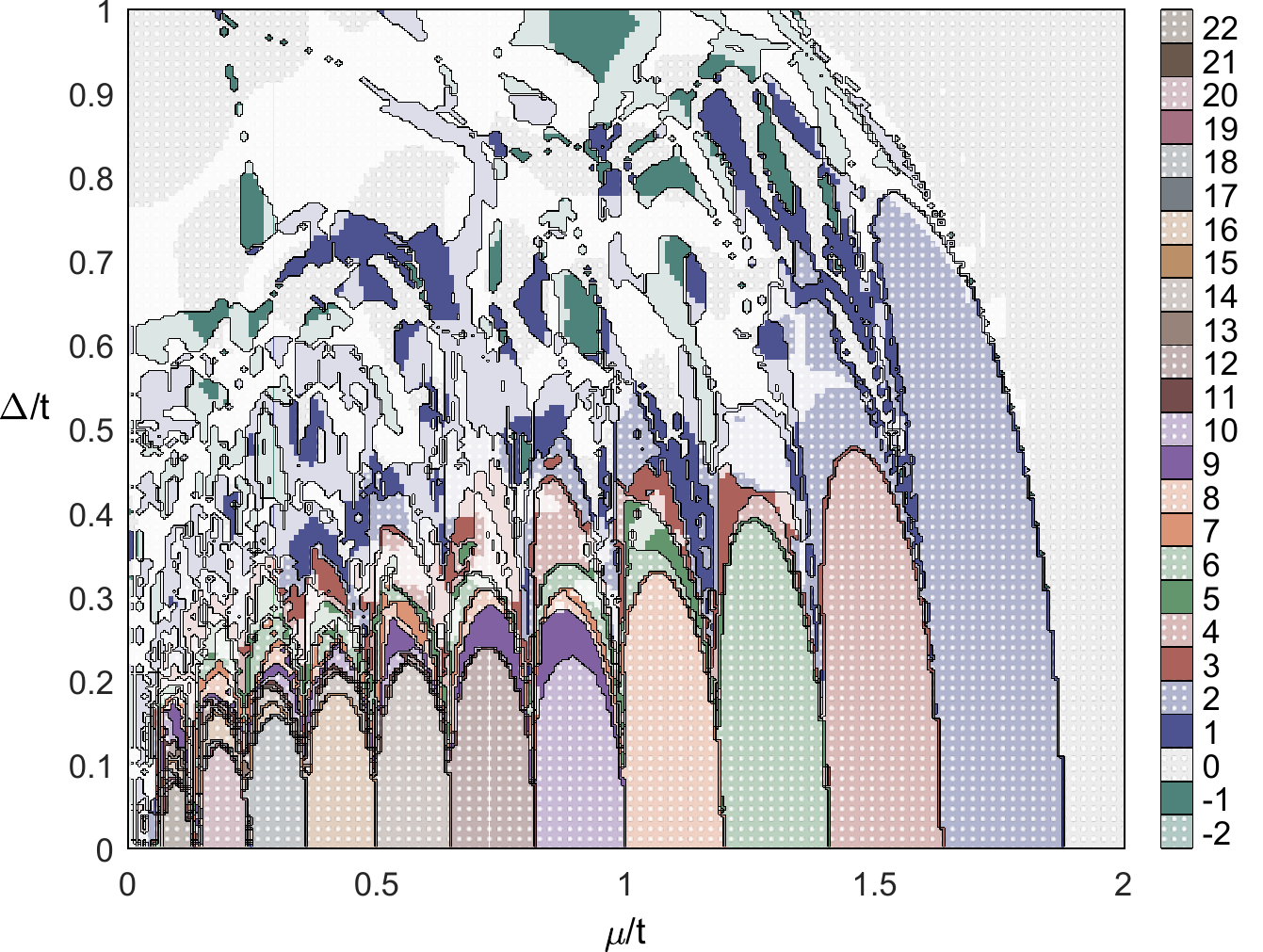}\\
(b) $f=1/3$
\caption{\label{fig:other_fluxes_phase_diagram}
Chern number phase diagrams as a function of the chemical potential $\mu$ and the induced SC pairing potential $\Delta$, for flux per SC plaquette (a) $f=\frac{2}{5}$ and (b) $f=\frac{1}{3}$. Both diagrams share their generic features with these of Fig.~\ref{fig:chern_phase_diagram} which was obtained for $f=\frac{1}{2}$. Substantial areas of odd Chern number, indicating a topological SC phase, can be seen at intermediate values of $\Delta$.
Regions of small energy gap (smaller than the bandwidth $t$ divided by the total number of sites) are shaded.
For display purposes, we used linear interpolation between the sampled points.
}
\end{figure}
\subsection{Robustness to flux}\label{subsec:flux_robustness}

We now turn to test the dependence of the above results on the magnetic flux threaded through the system.
Up to this point the flux was $f=\frac{1}{2}$ flux quantum per SC plaquette; we shall now examine two additional test cases, $f=\frac{2}{5}$ and $f=\frac{1}{3}$ (corresponding to $g=\frac{1}{20}$ and $g=\frac{1}{24}$ respectively), which give rise to simple enough unit cells that allow numerical investigations.
The Chern number phase diagrams for these two values of the flux are shown in Fig.~\ref{fig:other_fluxes_phase_diagram}.

These phase diagrams are similar to the one obtained for $f=\frac{1}{2}$ (see Fig.~\ref{fig:chern_phase_diagram}).
The effect of small $\Delta$ -- splitting of the two-fold transitions -- persists.
The effect of very large $\Delta$ -- destroying the topological phase -- persists as well.
The specifics do depend on the flux; in particular, the number of different phases at $\Delta=0$ varies.
Nevertheless, the general behavior appears to be flux-independent.

Remarkably, it seems that regardless of $f$, substantial areas of odd Chern number are formed in the parameter space at intermediate values of $\Delta$.
We have shown this behavior for three values of $f$.
The generic pattern we observed suggests that by tuning $\mu$ and $\Delta$, a topological phase can be stabilized for various values of $f$.

\section{Universality of the phase diagram}
\label{Sec:GenericForm}

Having calculated the phase diagrams in the previous section, we next wish to understand their generic properties.
Some details vary between different values of $f$, for example the shape of the boundary between different topological phases, the value of the gaps and the available Chern numbers of the phase diagram are not identical (see Sec.~\ref{Sec:SysAnalysis} and Figs.~\ref{fig:chern_phase_diagram},~\ref{fig:other_fluxes_phase_diagram}).
However, we find three features appearing in all cases:
(i) For large pairing potential $\Delta$ the system always becomes trivial.
(ii) For $\Delta=0$ the Chern numbers are even; as we change $\mu$, two-fold jumps of the Chern number occur at the transitions between the phases.
(iii) As we increase $\Delta$ for fixed $\mu$, successive transitions occur with mostly a reduction, and sometimes an increase, of the Chern number by one in each transition, until we reach the trivial phase with zero Chern number.

Motivated by these observations, we argue in this section that these features are universal and are shared also by other models with staggered $\Delta$.
To demonstrate this, we analyze simplified models in both one and two dimensions that can be treated semi-analytically.

We first show that when turning off the magnetic flux in the islands model, the system becomes topologically trivial for large $\Delta$.
We attribute this feature to localization of the Cooper pairs in regions with large pairing potential.
We show that in a simplified model of the 2D system that includes the magnetic flux, where the SC phase configuration can be approximately derived, the universal features appear in the phase diagram.

The observation regarding localization of Cooper pairs applies also in one dimension, as we show analytically.
This is demonstrated by analyzing a generalized one-dimensional Kitaev model with staggered pairing potential.
To obtain a system with large Chern numbers in 1D, we cascade several of these staggered Kitaev chains.
The resulting system is still one dimensional, as the number of chains is kept finite.
The system belongs to symmetry class BDI, which means it possesses a $\mathbb{Z}$ topological invariant~\cite{kitaev_periodic_2009, altland_nonstandard_1997, schnyder_classification_2008,fulga_scattering_2012}, just like the two-dimensional class D system we studied in Sec.~\ref{Sec:SysAnalysis}.
We use this analogy to engineer a system with the same pattern of topological transitions as the full 2D system, by judiciously controlling the coupling between the Kitaev chains.

\subsection{The effect of staggered SC}\label{subsec:effect-of-staggering}

We consider a generalization of Kitaev's chain model~\cite{kitaev_unpaired_2001}, which is described by the Hamiltonian
\begin{equation}
\begin{aligned}
H_{\mathrm{K}} &= -\mu \sum_{j=1}^N c^{\dagger}_j c_j \\
&+\sum_{j=1}^{N-1} \left( -t c^{\dagger}_j c_{j+1} + \Delta_j c_j c_{j+1} + \text{h.c.} \right),
\end{aligned}
\end{equation}
where $c^{\dagger}_j$ creates a single-species fermion at the $j^{\mathrm{th}}$ site of the 1D lattice, $\mu$ is the chemical potential, $t$ is the (real) nearest-neighbor hopping amplitude, and $\Delta_j$ is the SC $p$-wave pairing potential between sites $j$ and $j+1$.
When $\Delta_j$ is uniform, i.e. $\Delta_j=\Delta$ for all $j$, the model \cite{kitaev_unpaired_2001} supports a topological phase with Majorana end modes for $|\mu| < 2t$ for any $\Delta\neq 0$.

\begin{figure}
\begin{centering}
\includegraphics[width=0.9\linewidth]{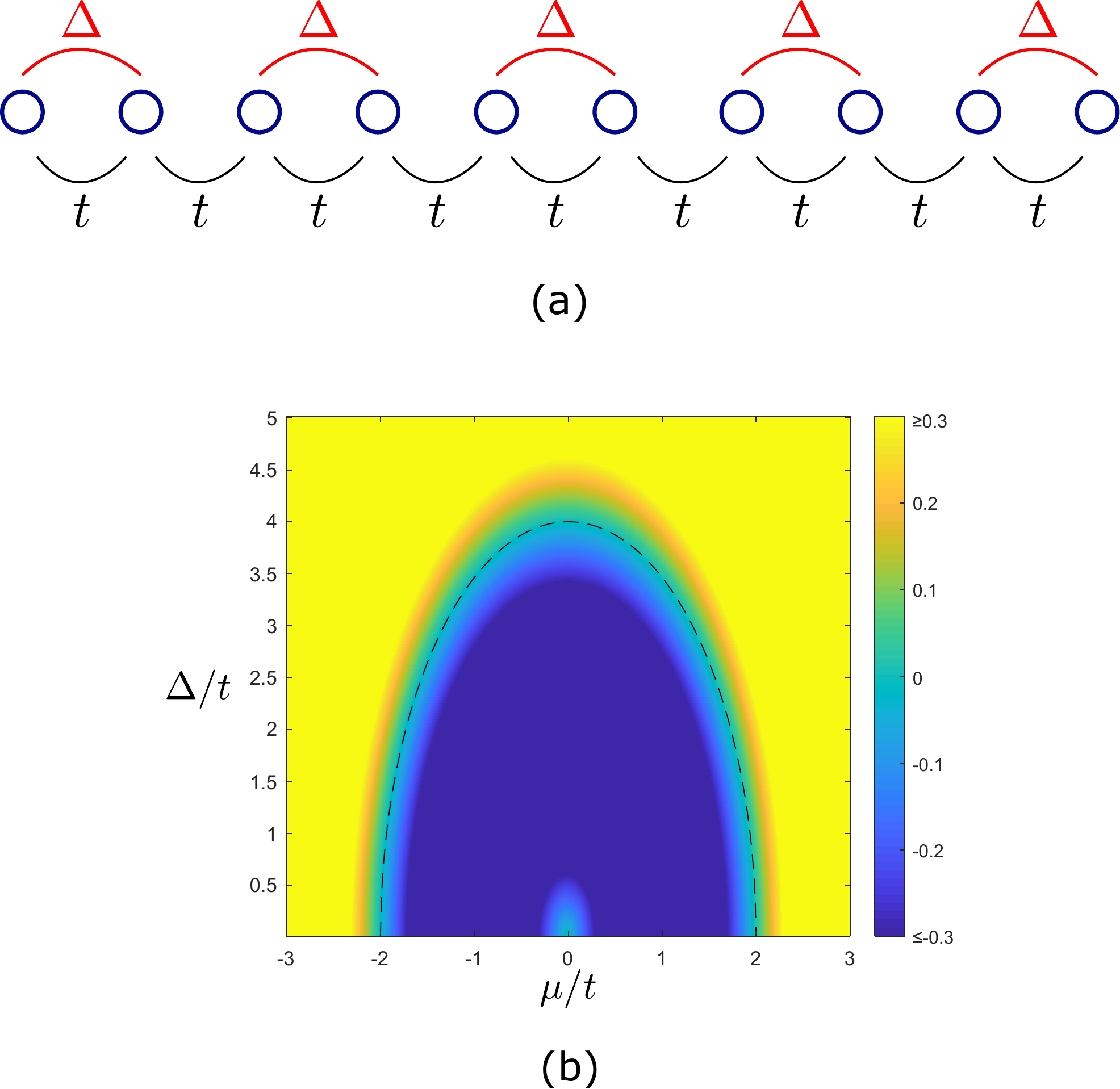}\caption{\label{fig:effect_of_staggered_SC_1d}
The effect of SC staggering on the topological phase diagram of the 1D Kitaev chain model.
(a) Illustration of the model: hopping exists between all nearest neighbors, and pairing only exists in every second bond.
(b) The phase diagram shows the sign of the Pfaffian (which equals $+1$ for the trivial phase and $-1$ for a topological phase, with a Chern number $+1$ or $-1$) multiplied by the energy gap as a function of the chemical potential, $\mu$ and the SC pairing potential $\Delta$.
The phase boundary is marked in a black dashed line for clarity, and we set a cutoff scale for the energy gap so the important details are clear.
The staggering of the SC limits the range of $\Delta$ in which the system is in the topological phase.
}
\par\end{centering}
\end{figure}

Consider now a staggered SC version of this model, where the pairing potential takes the values $\Delta_j= \Delta $ for odd $j$ and $\Delta_j=0$ for even $j$, as illustrated in Fig. \hyperref[fig:effect_of_staggered_SC_1d]{\ref{fig:effect_of_staggered_SC_1d}(a)}.
In momentum space, this modulation corresponds to an enlarged unit cell:
\begin{equation}
    H_{\text{st}}\left(k\right)=\begin{pmatrix}H_{\text{st}}^{(0)}\left(k\right) & H_{\text{st}}^{(\text{SC})}\left(k\right)\\
H_{\text{st}}^{(\text{SC})\dagger}\left(k\right) & -H_{\text{st}}^{(0)}\left(k\right)
\end{pmatrix}
\label{eq:staggered_Kitaev}
\end{equation}
with
\begin{subequations}
\begin{equation}
    H_{\text{st}}^{\left(0\right)}\left(k\right)=\begin{pmatrix}-\mu & t\left(1+e^{-ik}\right)\\
t\left(1+e^{ik}\right) & -\mu
\end{pmatrix},
\end{equation}
\begin{equation}
H_{\text{st}}^{\left(\text{SC}\right)}\left(k\right)= \frac{1}{2} \begin{pmatrix}0 & \Delta\\
-\Delta & 0
\end{pmatrix}
\end{equation}
\end{subequations}
(we take the lattice constant to be unity). By analyzing the Pfaffian~\cite{kitaev_unpaired_2001} of the Hamitlonian
$H_{\text{st}}\left(k\right)$
at $k=0,\pi$ we find that the topological phase appears for
$ \mu^2 + \left( \frac{\Delta}{2} \right)^2 < (2t)^2 $,
see the ellipse in Fig. \hyperref[fig:effect_of_staggered_SC_1d]{\ref{fig:effect_of_staggered_SC_1d}(b)}.
This means that SC staggering limits the range of $\Delta$ values for which the topological phase survives, reminiscent of the result obtained for the original 2D system.
Indeed, this staggered SC configuration can be thought of as the 1D analog of the SC islands configuration we studied in the 2D case.
The above result is in sharp contrast with the uniform SC case, where the phase boundaries do not depend on $\Delta$, provided it is non-zero.

\begin{figure}
\begin{centering}
\includegraphics[width=0.95\linewidth]{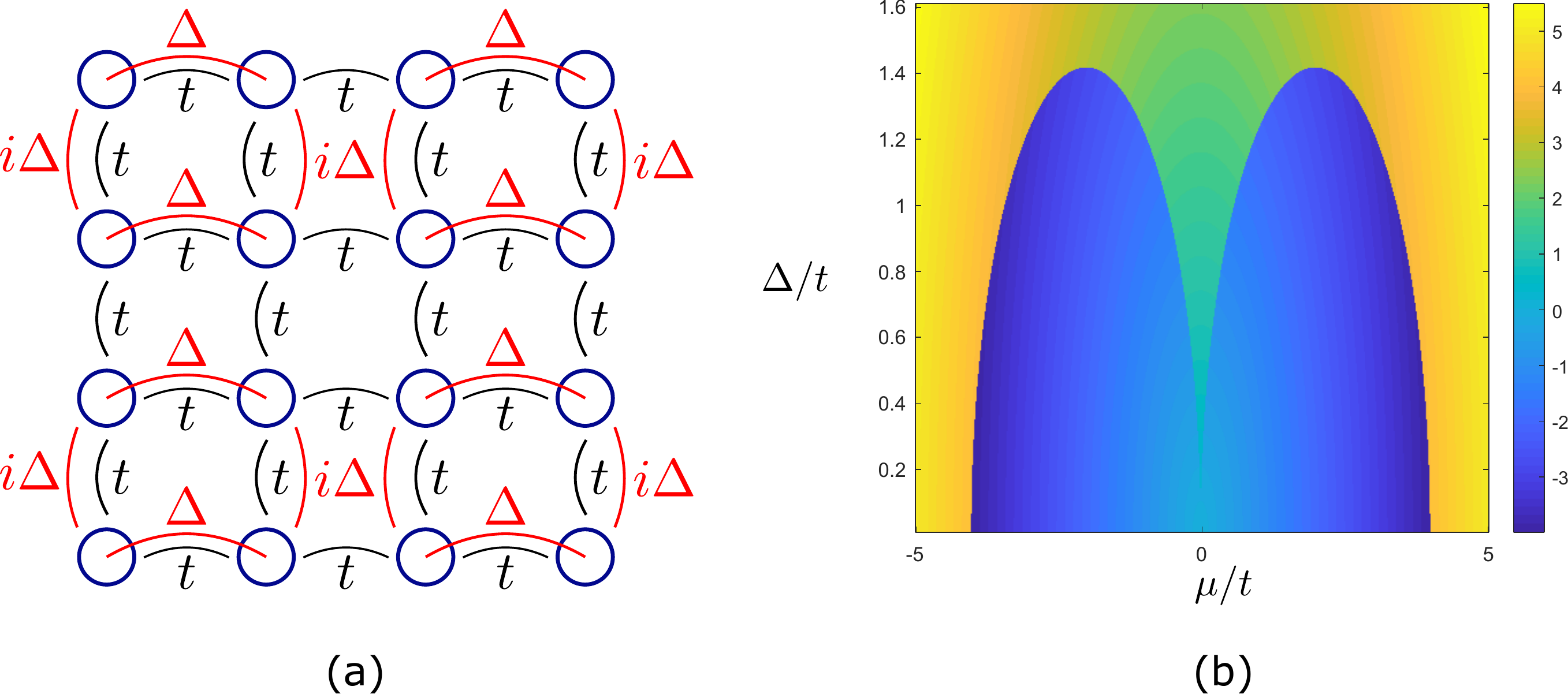}\caption{\label{fig:effect_of_staggered_SC_2d}
The effect of SC staggering on the topological phase diagrams of the 2D $p_x+ip_y$ SC.
(a) Illustration of the model: hopping exists between all nearest neighbors, and pairing only exists in islands.
(b) The phase diagram shows the sign of the Pfaffian (which equals $+1$ for the trivial phase and $-1$ for the topological phase) multiplied by the energy gap, as a function of the chemical potential $\mu$ and the SC pairing potential $\Delta$.
As in the 1D case (see Fig.~\ref{fig:effect_of_staggered_SC_1d}), the staggering of the SC limits the range of $\Delta$ in which the system is in the topological phase.
}
\par\end{centering}
\end{figure}

In order to further establish the effect of a staggered SC, we also studied a simple 2D model where a similar phenomenon occurs.
Consider the $p_x+ip_y$ model on a square lattice where the SC only exists in islands as illustrated in Fig. \hyperref[fig:effect_of_staggered_SC_2d]{\ref{fig:effect_of_staggered_SC_2d}(a)}.
This is a simplification of the model we started with to the case of zero magnetic flux.
For simplicity we choose each island to contain four sites (two along the $x$ direction and two along the $y$ direction).
The topological properties of the system may then be analyzed by the Pfaffian of the Hamiltonian at the time-reversal invariant momenta
$(k_x,k_y)=(0,0),(0,\pi),(\pi,0),(\pi,\pi)$.
The phase diagram for this model is shown in Fig. \hyperref[fig:effect_of_staggered_SC_2d]{\ref{fig:effect_of_staggered_SC_2d}(b)}, and it exhibits similar features to those of the 1D model.
In particular, the range of $\Delta$ supporting the topological phase is limited, unlike the uniform $p_x+ip_y$ case.
We have further verified that in order to get this effect, it is enough for $\Delta$ to be staggered in any way -- islands, stripes, or any other staggered patterns.

\subsection{Semi-analytical 2D model}\label{subsec:Semi-analytical-2D}

\begin{figure}[t]
\begin{centering}
\def\svgwidth{0.9\linewidth}
{\small %% Creator: Inkscape inkscape 0.92.4, www.inkscape.org
%% Accompanies image file 'stripes_schematic.pdf' (pdf, eps, ps)
\begingroup%
  \makeatletter%
  \providecommand\color[2][]{%
    \errmessage{(Inkscape) Color is used for the text in Inkscape, but the package 'color.sty' is not loaded}%
    \renewcommand\color[2][]{}%
  }%
  \providecommand\transparent[1]{%
    \errmessage{(Inkscape) Transparency is used (non-zero) for the text in Inkscape, but the package 'transparent.sty' is not loaded}%
    \renewcommand\transparent[1]{}%
  }%
  \providecommand\rotatebox[2]{#2}%
  \newcommand*\fsize{\dimexpr\f@size pt\relax}%
  \newcommand*\lineheight[1]{\fontsize{\fsize}{#1\fsize}\selectfont}%
  \ifx\svgwidth\undefined%
    \setlength{\unitlength}{1575.43754998bp}%
    \ifx\svgscale\undefined%
      \relax%
    \else%
      \setlength{\unitlength}{\unitlength * \real{\svgscale}}%
    \fi%
  \else%
    \setlength{\unitlength}{\svgwidth}%
  \fi%
  \global\let\svgwidth\undefined%
  \global\let\svgscale\undefined%
  \makeatother%
  \begin{picture}(1,0.53663239)%
    \lineheight{1}%
    \setlength\tabcolsep{0pt}%
    \put(0,0){\includegraphics[width=\unitlength,page=1]{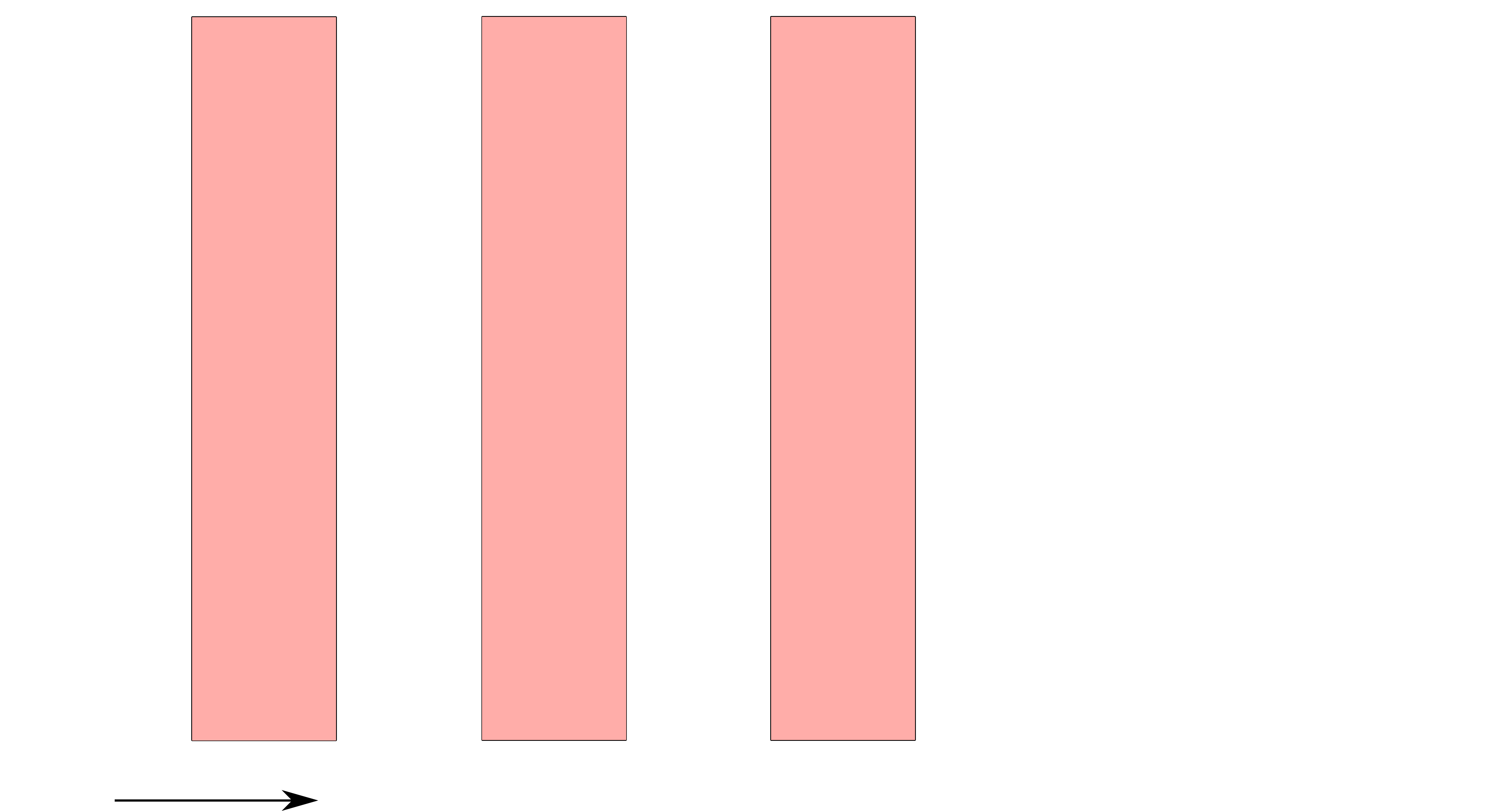}}%
    \put(0.23001678,0.00227128){\color[rgb]{0,0,0}\makebox(0,0)[lt]{\lineheight{1.25}\smash{\begin{tabular}[t]{l}$x$\end{tabular}}}}%
    \put(0.03093375,0.11674761){\color[rgb]{0,0,0}\makebox(0,0)[lt]{\lineheight{1.25}\smash{\begin{tabular}[t]{l}$y$\end{tabular}}}}%
    \put(0,0){\includegraphics[width=\unitlength,page=2]{stripes_schematic.pdf}}%
    \put(0.66820032,0.23136024){\color[rgb]{0,0,0}\makebox(0,0)[lt]{\begin{minipage}{0.34169799\unitlength}\centering \end{minipage}}}%
    \put(0,0.33856736){\color[rgb]{0,0,0}\makebox(0,0)[lt]{\begin{minipage}{0.09040759\unitlength}\centering $\vec{B}$\end{minipage}}}%
    \put(0,0){\includegraphics[width=\unitlength,page=3]{stripes_schematic.pdf}}%
  \end{picture}%
\endgroup%
}
\par\end{centering}
\caption{\label{fig:stripes_schematic}
Illustration of the stripes system (top view) described in Eq.~\eqref{eq:H_stripes}. Blue circles denote lattice sites in the tight-binding model, and $p$-wave superconductivity is introduced only on the red areas. A magnetic field $B$ is applied perpendicular to the system's plane. The ground-state phase configuration of the SC in this model may be analytically approximated, see Eqs.~\eqref{eq:theta_SC_x}--\eqref{eq:theta_SC_y}.}
\end{figure}

Having established the importance of a staggered $\Delta$, we now introduce a 2D model with magnetic flux which can be understood almost entirely from analytical considerations.
We consider a system of coupled SC stripes, with a constant perpendicular magnetic field, see Fig.~\ref{fig:stripes_schematic}.
The virtue of this model is that in the strongly anisotropic limit, where the stripes can be treated independently, we can analytically determine the ground-state SC phases.

The Hamiltonian takes the form
\begin{equation}
\begin{aligned}H= & -\mu\sum_{m,n}c_{m,n}^{\dagger}c_{m,n} + \\
 & \sum_{m,n}\left[t_x c_{m,n}^{\dagger}c_{m+1,n}+t_y e^{i\theta_{\text{P}}(m,n)}c_{m,n}^{\dagger}c_{m,n+1}\right. \\
 & \left. +\Delta e^{i\theta_{\text{SC}}^x (m,n)}c_{m,n}c_{m+1,n} \cdot (m \text{ mod } 2)  \right.\\
 & \left. +i\Delta e^{i\theta_{\text{SC}}^y (m,n)}c_{m,n}c_{m,n+1} + \text{h.c.} \right].
\end{aligned}
\label{eq:H_stripes}
\end{equation}
The $j^{\text{th}}$ stripe lies at $m=2j-1,2j$.
The Peierls phase in the Landau gauge is
$\theta_{\text{P}}(m,n) = -2\pi \frac{\Phi}{\Phi_0} m$.
In the ground state, the gauge-invariant phase difference between every two SC bonds is zero~\cite{tinkham_introduction_2004}, yielding
\begin{equation}
\theta_{\text{SC}}(\vec{r}_b) - \theta_{\text{SC}}(\vec{r}_a) = \frac{4\pi}{\Phi_0} \int_{\vec{r}_a}^{\vec{r}_b} \vec{A}\cdot d\vec{r}.
\label{eq:SC_phase_difference}
\end{equation}
The $\theta_{\text{SC}}^x$ phases reside at half-integer $x$ and integer $y$ (they connect two adjacent sites along $x$), and vice versa for $\theta_{\text{SC}}^y$.
Therefore, in our gauge choice, a reasonable configuration is given by
\begin{subequations}
\begin{equation}
\theta_{\text{SC}}^x(m,n) = -4\pi \frac{\Phi}{\Phi_0}m n , \label{eq:theta_SC_x}
\end{equation}
\begin{equation}
\theta_{\text{SC}}^y(m,n) = -2\pi \frac{\Phi}{\Phi_0}m n. \label{eq:theta_SC_y}
\end{equation}
\end{subequations}

Unlike the more sophisticated model studied in Sec.~\ref{Sec:SysAnalysis}, here the SC phase configuration is known.
It is indeed an approximation -- each stripe is treated independently of the others -- but at least in the limit $t_y\gg t_x$ it is sensible.
Determining the phase configuration correctly is of paramount importance: The phase diagram, and in particular the topological regions, are extremely sensitive to the SC phases (this is also demonstrated in Appendix~\ref{appendix:additional_2d_results} for the full model).

\begin{figure}
\begin{centering}
\includegraphics[width=0.9\linewidth]{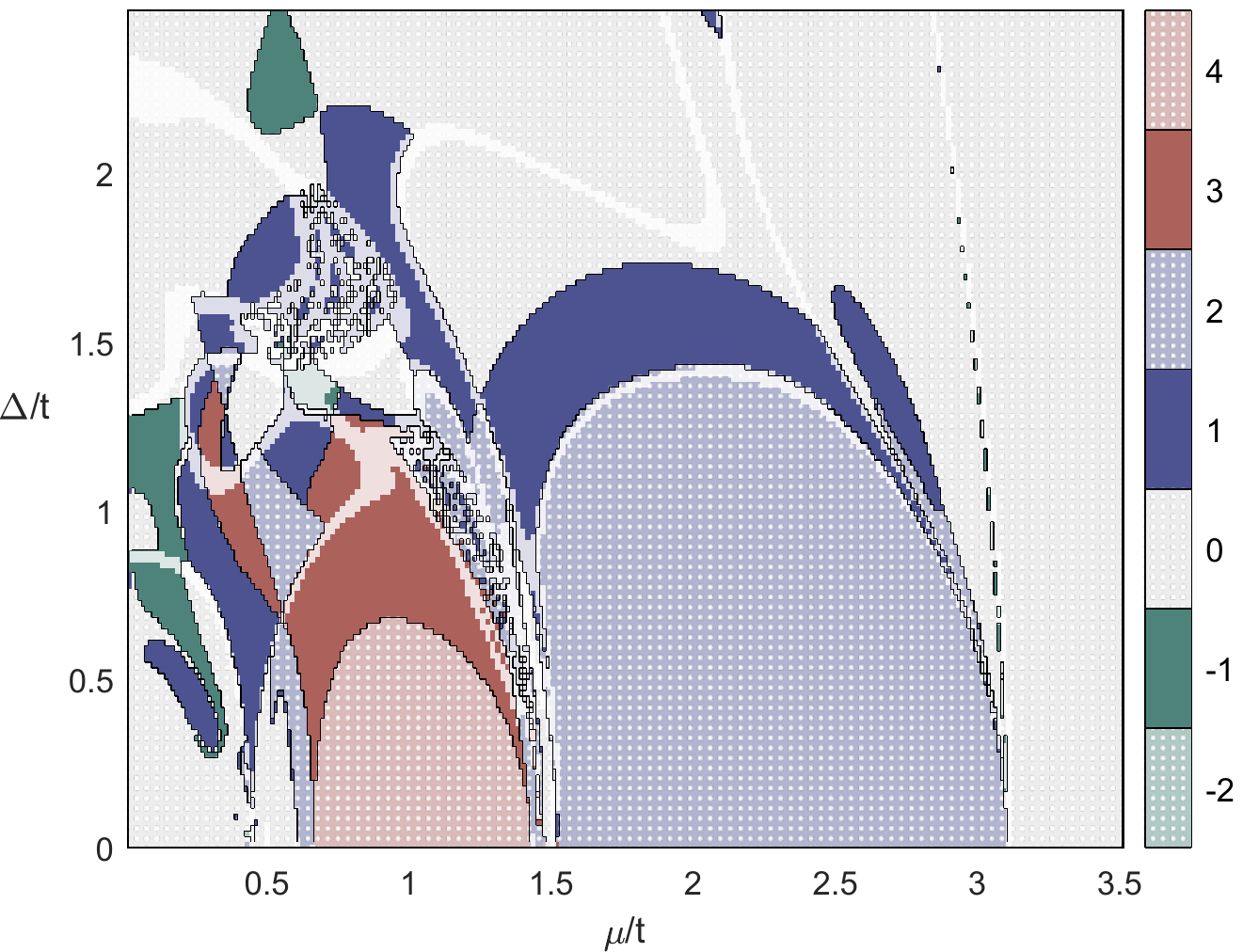}\caption{\label{fig:stripes_phase_diagram}
Topological phase diagram for stripes model Eq.~\eqref{eq:H_stripes}, as a function of the chemical potential $\mu$ and the pairing potential $\Delta$. The parameters used are $t_x=t_y=t=1$ and the flux is $\Phi/\Phi_0=1/6$.
This phase diagram exhibits universal features (cf. Figs.~\ref{fig:chern_phase_diagram},~\ref{fig:other_fluxes_phase_diagram} of the full numerical model): At $\Delta=0$ there are only even Chern numbers which gradually decrease as $\mu$ is swept;
at large $\Delta$ the system becomes topologically trivial;
intermediate $\Delta$ gives rise to topological phases with odd Chern numbers.
Regions of small energy gap (smaller than the bandwidth $t$ divided by the total number of sites) are shaded.
}
\par\end{centering}
\end{figure}

Armed with the phase configuration, we are in a position to analyze the Chern number, just as was done in Sec.~\ref{Sec:SysAnalysis}.
The topological phase diagram for this model, as a function of the chemical potential $\mu$ and the pairing potential $\Delta$, is shown in Fig.~\ref{fig:stripes_phase_diagram}.
The phase diagram shares the important features mentioned in Subsec.~\ref{subsec:Chern} with the phase diagrams of the full model (see Figs.~\ref{fig:chern_phase_diagram},~\ref{fig:other_fluxes_phase_diagram}).
In particular, it displays the crossover from the Hofstadter transitions at $\Delta=0$ to the localized superconducting state at large $\Delta$, passing through a series of topological phases at intermediate $\Delta$.

The stripes structure, chosen here for its simplicity, is not special.
We tested two additional variants of this semi-analytical 2D model: one where the SC stripes are replaced by SC islands, and one where the flux is only threaded in the normal regions between the SC stripes.
We have also corroborated the results by applying different magnetic fluxes.
The phase diagrams in all of these variants are similar, and most importantly they all share the features we refer to as universal.

\subsection{Stacking several staggered Kitaev chains}\label{subsec:Stacking-Kitaev-chains}

Let us now turn our attention to 1D.
Using the staggered Kitaev chain Eq.~\eqref{eq:staggered_Kitaev} with $\Delta=\mu$ as a building block, we present a way to construct a phase diagram similar to the ones of the full 2D model shown in Sec.~\ref{Sec:SysAnalysis}.
This is achieved by stacking $M$ staggered Kitaev chains, each having a different hopping parameter $t$.
In the absence of coupling between the chains, we get independent Chern number transitions at the $\mu$ values appropriate for each $t_j$ ($j=1,\ldots,M$) and $\Delta$.
Since we want to mimic the behavior of the full 2D system, we take an even number of chains and choose their hopping parameters to be pairwise equal, i.e.
$\left\{ t_1,t_1,t_2,t_2,\ldots, t_{M/2},t_{M/2} \right\}$.
This choice guarantees that in the absence of inter-chain coupling, the Chern number may only change by an even number at each transition.

Next, we introduce normal (non-SC) coupling of strength $w$ between neighboring chains.
The role of this coupling is similar to that of $\Delta$ in the 2D case:
when turned on, it splits the two-fold Chern number transitions into single transitions, where the Chern number changes by $\pm 1$.
At large $\mu$ or large $w$, the Chern number vanishes due to the SC staggering.
Qualitatively, we infer that this construction reproduces the universal features we mentioned before.

\begin{figure}
\begin{centering}
\includegraphics[width=0.9\linewidth]{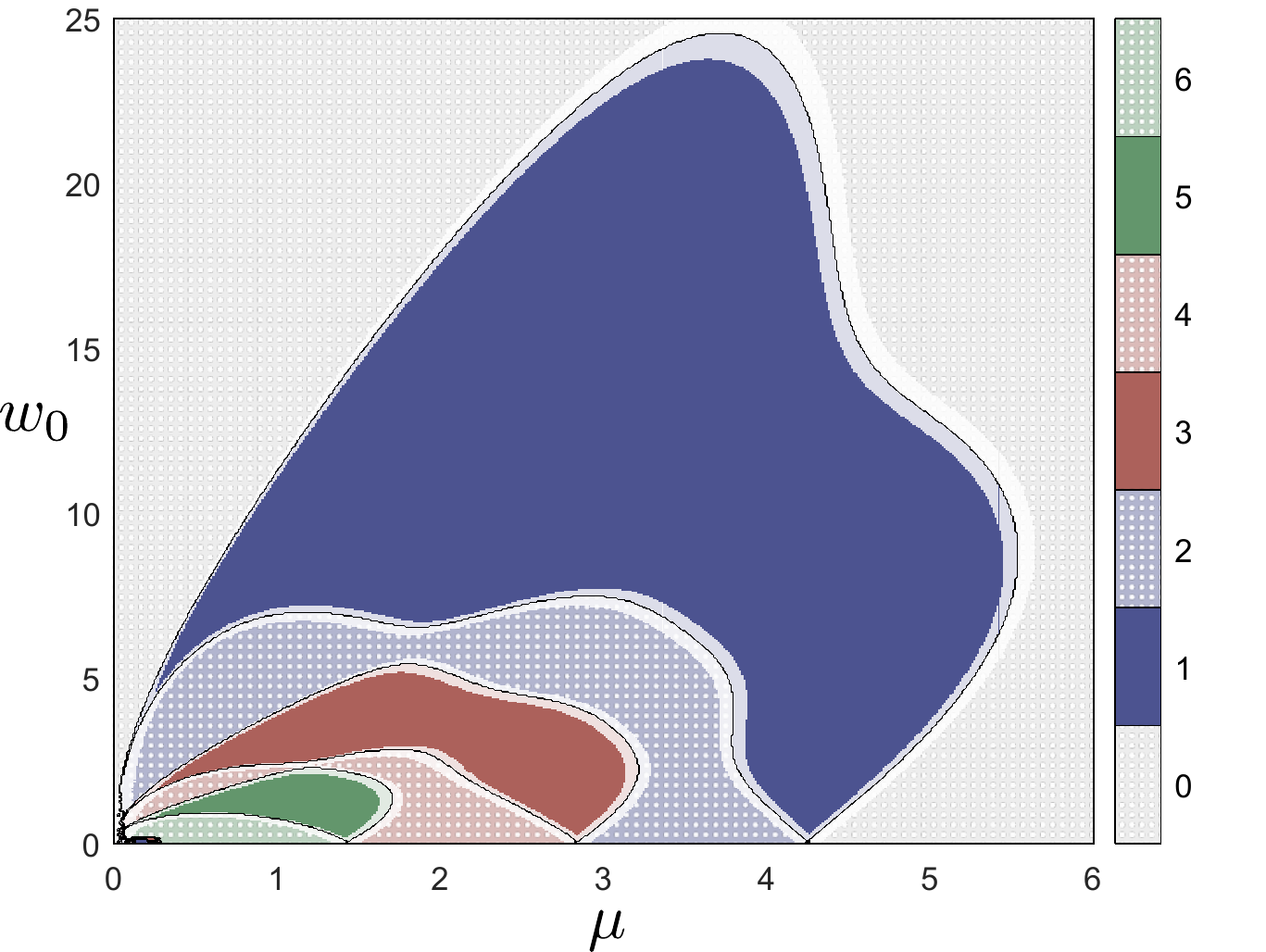}\caption{\label{fig:coupled_Kitaev_phase_diagram}
Topological phase diagram for the coupled staggered Kitaev chains, as a function of the chemical potential $\mu$ and the coupling strength $w_0$.
This specific system is composed of six coupled chains, with hopping parameters
$\vec{t}=\left\{ 1,1,2,2,3,3 \right\}$ and $\mu_0=0.01$.
When the chains are decoupled, i.e. $w_0=0$, the Chern number is even, since the hopping parameters were chosen to be pairwise equal.
Finite $w_0$ splits these transitions, much like $\Delta$ does in the full 2D case (cf. Figs.~\ref{fig:chern_phase_diagram},~\ref{fig:other_fluxes_phase_diagram},~\ref{fig:stripes_phase_diagram}).
Regions of small energy gap are shaded.}
\par\end{centering}
\end{figure}

In order to make the phase diagram of this model even more similar to that of the original 2D system, we perform a slight modification to $w$.
A single staggered chain with $\mu=\Delta$ undergoes a transition of the Chern number from $-1$ to $1$ at $\mu=0$.
Thus, our coupled system undergoes a transition from $-2M$ to $2M$ at $\mu=0$ without passing at zero Chern number.
Let us then choose
$w=w_0/\sqrt{\mu+\mu_0}$, where $w_0$ is the bare coupling strength and $\mu_0$ is a small constant.
This way, at $\mu\rightarrow 0$ the actual coupling $w$ becomes strong, thus driving the Chern number to zero.
This transformation, which can be seen as a redefinition of the axes, gives rise to the phase diagram shown in Fig.~\ref{fig:coupled_Kitaev_phase_diagram}.
The resulting phase diagram is similar to the 2D phase diagrams shown in Figs.~\ref{fig:chern_phase_diagram},~\ref{fig:other_fluxes_phase_diagram},~\ref{fig:stripes_phase_diagram}.
We therefore conclude that this simple 1D model captures most of the generic features of the complicated 2D models.
This observation supports the notion of universality in the phase diagrams, which applies to systems related by dimensionality and symmetry properties: class D in 2D and class BDI in 1D.

\section{Discussion}\label{Sec:Discussion}
In this paper, we studied the orbital effects of a magnetic field on the topological properties of the $p_x+ip_y$ SC.
Motivated by relevant experiments, we focused on a model of SC islands arranged in a square lattice.
After determining the ground state configuration of the SC, we derived the topological phase diagram for different values of the magnetic flux (see Figs.~\ref{fig:chern_phase_diagram},~\ref{fig:other_fluxes_phase_diagram}).

The tunable parameters in our model are the magnetic flux $\Phi$, the chemical potential $\mu$, and the strength of the induced SC pairing potential $\Delta$.
Our results suggest that for general values of the flux, it is possible to tune into a topological phase supporting Majorana edge modes by varying $\mu$ and $\Delta$.
The regions in $\mu,\Delta$ space supporting the topological phase are substantial, so the parameters do not have to be extremely fine-tuned.

Experimental detection of the Majorana edge modes can be done by interference \cite{stern_proposed_2006,bonderson_detecting_2006,nilsson_splitting_2008}.
In addition, these Majorana modes can be detected by measuring the heat conductivity, which is expected to be a half-integer multiple of
$\pi^2 k_B^2 T/3h$ in the presence of a chiral Majorana edge mode~\cite{banerjee_observation_2018}.
We also note that in the model we analyzed, the islands are connected by few sites, without any normal regions between them.
The universality of the phase diagram we found suggests that the inclusion of normal regions between the islands, which exist in several experimental realizations, may not alter the main features of the phase diagram.

Our model should be understood as an effective description of a more complicated physical system.
We assumed the existence of an induced $p$-wave pairing potential and single-species fermions.
One can take a more microscopic approach and study spinful electrons with spin-orbit coupling, proximity coupled to an $s$-wave SC with an applied Zeeman field, which can give rise to induced $p$-wave pairing.
It is also possible (at least numerically) to account for disorder, which is not expected to have a drastic effect on the results provided it is smaller than the energy gap.
Likewise, our zero-temperature study may be generalized to finite temperatures, but as long as the SC gap is larger than the temperature, our results are expected to hold qualitatively.

We also studied simplified, analytically solvable models in 1D and 2D which share common features with the original 2D model.
We found that the islands structure (and more generally the staggering of the SC pairing potential) has a profound implication on the topological phase diagram:
Pair localization at large $\Delta$ drives the system away from the topological phase.
We showed that a model consisting of coupled staggered 1D Kitaev-like chains exhibits a topological phase diagram which greatly resembles those of the original 2D model (see Fig.~\ref{fig:coupled_Kitaev_phase_diagram}).
These findings suggest that several features of the phase diagram are shared by many models.
The results shed light on those obtained for the original model, which relied mainly on numerics.

This notion of universality may also be viewed from a general mathematical perspective.
Consider a model with two tunable parameters $p_1$ and $p_2$, such that at $p_1=0$ the Chern number has to be even.
Assume further that at large $|p_1|$ and $|p_2|$ the Chern number has to vanish, and that at $p_1\neq0$ the generic behavior is one-fold Chern number transitions.
Under such settings, it is almost impossible to construct a phase diagram which is topologically distinct from those shown in Figs.~\ref{fig:chern_phase_diagram},~\ref{fig:other_fluxes_phase_diagram},~\ref{fig:stripes_phase_diagram},~\ref{fig:coupled_Kitaev_phase_diagram}.
We conclude that the phase diagrams we found have universal features and may appear under very general settings.

\section*{Acknowledgements}
We are grateful to Eran Sagi for participating in the early stages of this project.
We are thankful for discussions with B. I. Halperin.
The research was supported by the European Union's Horizon 2020 research and innovation
programme grant agreement LEGOTOP No 788715, the DFG (CRC/Transregio 183, EI 519/7- 1), the Israel Science Foundation (ISF) and the Binational Science Foundation (BSF).

\appendix
%\onecolumngrid

\section{\label{appendix:ground_state}Determining the Ground State}

The Hamiltonian Eq.~\eqref{eq:islands_hamiltonian} depends on all the islands' phases $\{\theta_{j}\}$.
Here we describe a direct numerical method of finding the ground state configuration of $\{\theta_{j}\}$ .

The system possesses a global $U(1)$ symmetry, i.e., the transformation
$\theta_j \longmapsto \theta_j + \gamma$
for all $j$ leaves the energy invariant for constant $\gamma$.
Therefore, without loss of generality we set one of the phases, say $\theta_1$, to zero.
The Hamiltonian thus depends on $N_{\text{islands}}-1$ phases.

When writing the Hamiltonian in BdG form as in Eq.~\eqref{eq:H_BdG}, the many-body ground state energy is given by the sum of all negative eigenvalues of $H$, i.e.
\begin{equation}
E_{\text{MB}} = \sum_{n=1}^{2 N_x N_y} E_n \Theta\left(-E_n\right),
\end{equation}
where $\Theta(x)$ is the step function and $E_n$ are the eigenvalues of $H$.
$E_{\text{MB}}$ is a function of $\{\theta_{j}\}$ , and our goal is to find the configuration of $\{\theta_{j}\}$  that minimizes it.

We carried out the optimization in real space, using a finite system with periodic boundary conditions.
We used the optimization algorithm ``interior-point"~\cite{karmarkar_new_1984,forsgren_interior_2002}.
In order to increase the probability of finding the global minimum point, we used many random initial conditions for $\{\theta_{j}\}$.
We compared this numerical method and the known frustrated XY ground states for several fluxes, and found that the configurations are qualitatively similar.
We note that the optimization is sensitive to the initial conditions, and many local minima exist, so multiple runs are necessary in order to get a good agreement with the results of the frustrated XY model.
The numerical optimization shows that the ground-state configurations of the frustrated XY model are at least local minima of our model.

\section{\label{appendix:geometry}Geometric relation between $f$ and $g$}

\begin{figure}
\def\svgwidth{0.9\linewidth}
{\small 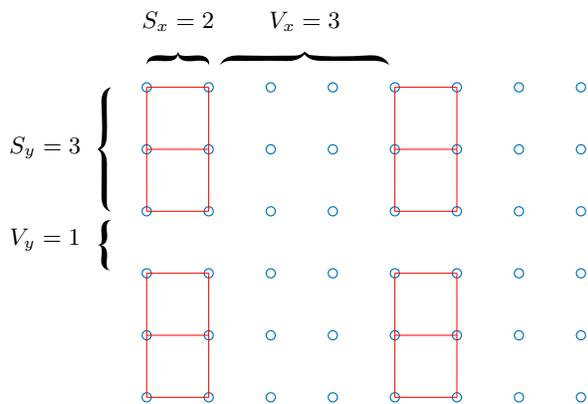}
\caption{\label{fig:geometry_SV}
Geometrical meaning of the $S_i,V_i$ parameters.
In this example there are 4 islands, each containing $S_x=2$ lattice sites along the $x$ direction and $S_y=3$ sites along the $y$ direction.
The different islands are separated by $V_x=3$ normal plaquettes along the $x$ direction and $V_y=1$ normal plaquettes along the $y$ direction.
}
\end{figure}

Let us denote the number of lattice sites per dimension per island by $S_{i}$ where $i=x,y$ is the spatial dimension, and the number of plaquettes separating adjacent islands per dimension by $V_{i}$ (see Fig.~\ref{fig:geometry_SV}).
Then, the area of a superlattice unit cell is  $A_{\text{SL}}=\left(S_{x}+V_{x}-1\right)\left(S_{y}+V_{y}-1\right)$,
and the area of a magnetic unit cell is $A_{\text{M}}=\left(S_{x}-1\right)\left(S_{y}-1\right)$.
The flux per magnetic plaquette is $g=\frac{\Phi}{h/e}$ and the flux per superlattice unit cell is $f=\frac{\Phi_{\text{SC}}}{h/2e}$ (see Fig.~\ref{fig:islands_system}).
The factor of $2e$ comes from the fact that the relevant flux quantum for the superconductor is that of Cooper pairs.
The relation between $f$ and $g$ is thus
\begin{equation}
f=2g\cdot\frac{A_{\text{SL}}}{A_{\text{M}}}=2g\cdot\frac{\left(S_{x}+V_{x}-1\right)\left(S_{y}+V_{y}-1\right)}{\left(S_{x}-1\right)\left(S_{y}-1\right)}\ .
\end{equation}
In the current study we used $S_{x}=S_{y}=2$ and $V_{x}=V_{y}=1$, yielding the relation $f=8g$.
The largest value of $f$ we investigated is $\frac{1}{2}$ and therefore the largest value of $g$ is $\frac{1}{16}$, so the approximation of a constant phase inside each island is reasonable.

\section{\label{appendix:chern_realspace}Real space calculation of the Chern number}
Here we review the method devised in Refs.~\cite{yi-fu_coupling-matrix_2013,loring_disordered_2011} of calculating the Chern number from the real space Hamiltonian.
Consider a 2D lattice with $N=L_x L_y$ unit cells, and denote their positions by $\vec{r}=(x,y)$ where $x,y$ are integers. Let us define twisted periodic boundary conditions by
\begin{equation}
\begin{gathered}
\varphi_{\theta}(x+L_x,y) = e^{i\theta_x} \varphi_{\theta}(x,y), \\
\varphi_{\theta}(x,y+L_y) = e^{i\theta_y} \varphi_{\theta}(x,y), \\
\theta=(\theta_x,\theta_y),
\end{gathered}
\end{equation}
where
$\varphi_{\theta}^{m} (x,y)$
are the single-particle wavefunctions ($m=0,\ldots,M-1$ where $M$ is the number of electrons), which are vectors in the space
of inner degrees of freedom.
The many-body ground-state wavefunction
$\Psi_{\theta}(\left\{\vec{r}_i \right\})$
is the Slater determinant of
the single-particle wavefunctions of the occupied states, and the Chern number is given by
\begin{equation}
C=\frac{1}{2\pi i}\int_{T_{\theta}} d\theta \bra{\nabla_{\theta}\Psi_{\theta}} \times \ket{\nabla_{\theta}\Psi_{\theta}},
\end{equation}
where $T_{\theta}$ is the torus
$0 \leq \theta_x,\theta_y \leq 2\pi$.

Going to momentum space, we denote by
$F^{m}_{\theta} (\vec{k})$
the Fourier components of $\varphi_{\theta}^{m} (x,y)$.
The twisted boundary conditions dictate the allowed momenta,
\begin{equation}
\begin{gathered}
\vec{k} = \vec{k}^{(0)} + \vec{q}, \\
\vec{k}^{(0)} = \left( \frac{2\pi n}{L_x}, \frac{2\pi m}{L_y} \right) \quad \text{where } \, n,m\in\mathbb{Z}, \\
\vec{q} = \left( \frac{\theta_x}{L_x},\frac{\theta_y}{L_y}\right) .
\end{gathered}
\end{equation}

Denoting
$F^{m}_{q}(\vec{k}^{(0)}) \equiv F^{m}_{\theta} (\vec{k})$,
we note that the many-body wavefunction in momentum space
$\Phi_{q}\left(\left\{ \vec{k}_i^{(0)} \right\}\right)$
is just the Slater determinant of
$F^{m}_{q}(\vec{k}^{(0)})$.
Thus, upon substituting
$\partial_{\theta_{\mu}}=L_{\mu}^{-1}\partial_{q_{\mu}}$ (here $\mu=x,y$) we obtain an equation for the Chern number:
\begin{equation}
C=\frac{1}{2\pi i}\int_{R_q} dq \bra{\nabla_{q}\Phi_{q}} \times \ket{\nabla_{q}\Phi_{q}},
\end{equation}
where $R_q$ is the rectangle
$\left[0,\frac{2\pi}{L_x}\right] \times \left[0,\frac{2\pi}{L_y}\right]$.
Using Stokes' theorem, the above expression can be written as a winding number along the boundary $\partial_{R_q}$,
\begin{equation}
C = \frac{1}{2\pi i} \oint_{\partial_{R_q}} d\vec{\ell_q} \cdot \braket{\Phi_q}{\nabla_q \Phi_q}.
\end{equation}

Next, we divide $\partial_{R_q}$ into $N_q$ small line segments and replace the derivatives and integral by their discrete counterparts to obtain
\begin{equation}
C = \frac{1}{2\pi} \sum_{\alpha=0}^{N_q} \arg \left[ \det \left( C_{\alpha,\alpha+1} \right) \right],
\end{equation}
where
$C_{\alpha,\alpha+1}^{m,n} = \braket{F^{m}_{q_{\alpha}}}{F^{n}_{q_{\alpha+1}}}$
are the coupling matrices and $q_{\alpha}$ are the endpoints of the small line segments.

If $L_x,L_y \gg 1$, i.e., we consider a sufficiently large system, it is enough to take $N_q=4$, corresponding to $\theta=0$ (i.e. just the four corners of $\partial_{R_q}$).
In this case we only need to deal with quantities calculated for periodic boundary conditions.
Transforming the coupling matrices back to real space for $\theta=0$ we obtain
\begin{equation}
C^{m,n}_{\alpha,\alpha+1} = \bra{\varphi_{\theta=0}^{m}} e^{i(\vec{q}_{\alpha}-\vec{q}_{\alpha+1})\cdot\vec{r}} \ket{\varphi_{\theta=0}^{n}}.
\end{equation}
We may thus define the matrix
$\tilde{C} = C_{0,1} C_{1,2} C_{2,3} C_{3,0}$,
diagonalize it to obtain the eigenvalues
$\left\{ \lambda_m \right\}$,
and calculate the Chern number:
\begin{equation}
C = \frac{1}{2\pi} \sum_{m=0}^{M-1} \arg{\lambda_m}.
\end{equation}
If the Hamiltonian is written in the BdG form, this method yields the BdG Chern number $\mathcal{N}$.

\section{\label{appendix:additional_2d_results}Additional numerical results in the 2D system}
We now present several additional results in the full 2D system, described by the Hamiltonian Eq.~\eqref{eq:H_BdG}.
These results were obtained numerically, and they support the arguments given in Sec.~\ref{Sec:SysAnalysis} of the main text: the vanishing of the energy gap near $\mu=0$ and the strong dependence of the topological phase diagram on the SC phases.

\begin{figure}[ht]
\includegraphics[width=0.95\linewidth]{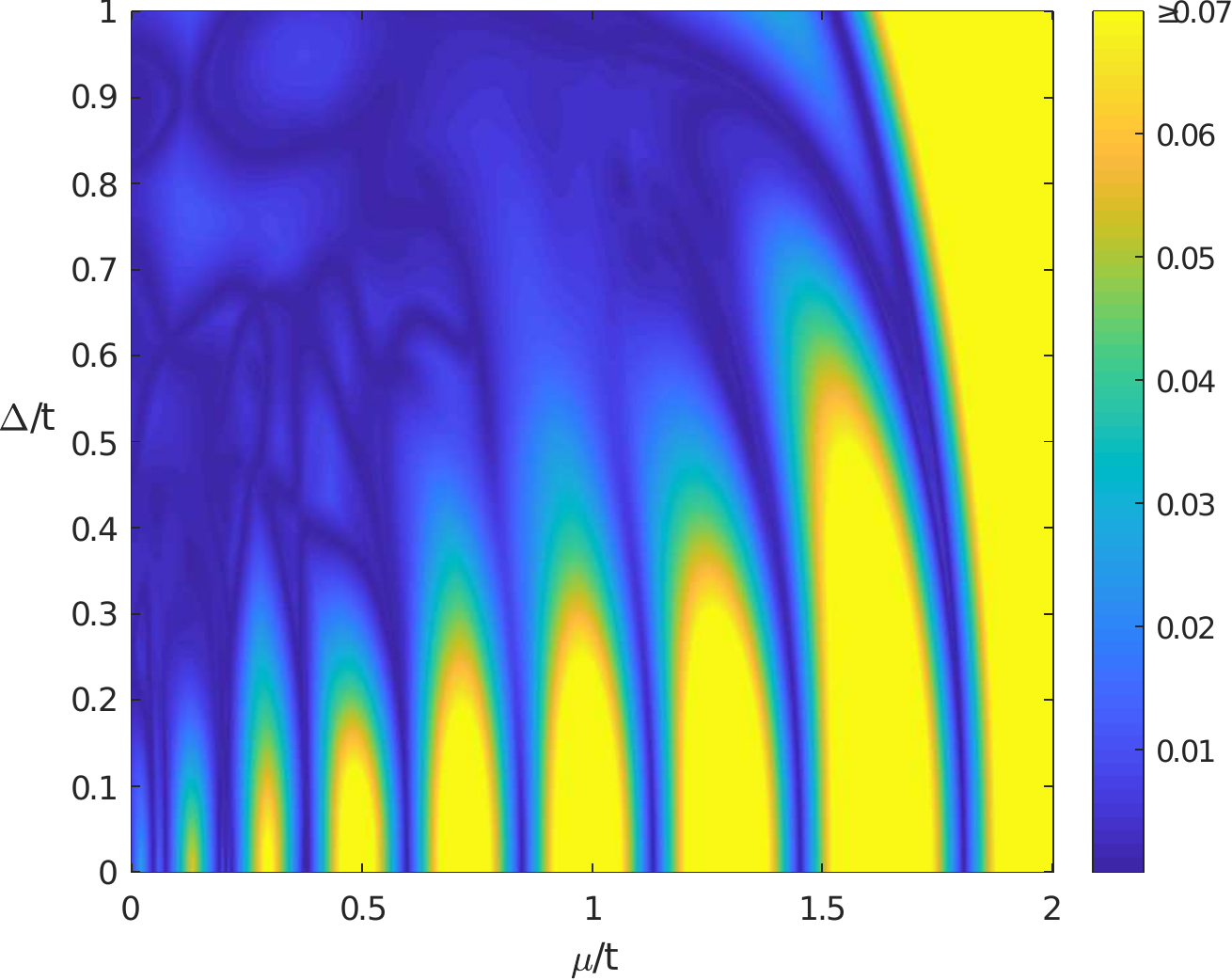}
\caption{\label{fig:energy_gap_map}
Energy gap map (in units of the hopping amplitude $t$) for $f=\frac{1}{2}$, as a function of the chemical potential $\mu$ and the SC pairing potential $\Delta$.
To enable clear distinction between small values, we set a cutoff scale at $0.07t$.
The ``domes", which are observed in the phase diagram Fig.~\ref{fig:chern_phase_diagram} as well, are gapped whereas the phases near $\mu=0$ are gapless (or posses a small energy gap).
}
\end{figure}

Fig.~\ref{fig:energy_gap_map} shows the energy gap as a function of the model's parameters for $f=\frac{1}{2}$.
Compared with the phase diagram Fig.~\ref{fig:chern_phase_diagram}, it is evident that the topological phases are gapped.
In addition, this map makes the transitions of the Chern number transparent -- they are accompanied by a closing of the energy gap.

\begin{figure}
\includegraphics[width=0.95\linewidth]{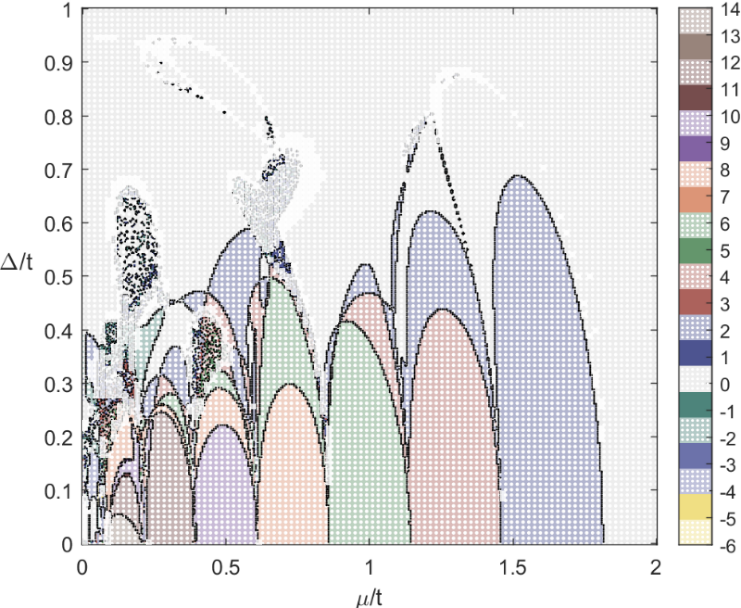}
\caption{\label{fig:aligned_phase_diagram}
Topological phase diagram for $f=\frac{1}{2}$ flux quantum per superconducting plaquette, as a function of the chemical potential $\mu$ and the induced SC pairing potential $\Delta$, for the case where the SC phases are uniform.
The phase diagram has the same ``domes" structure of the ground-state phase diagram (see Fig.~\ref{fig:chern_phase_diagram}), but it almost does not support odd Chern number phases.
}
\end{figure}

Fig.~\ref{fig:aligned_phase_diagram} shows the Chern number phase diagram for $f=\frac{1}{2}$, in the case where the SC phases are uniform, i.e. $\theta_j=0$ for all $j$.
This phase configuration is not the ground state due to the presence of the magnetic field.
Though it resembles the true phase diagram Fig.~\ref{fig:chern_phase_diagram} in its general shape, this phase diagram does not support odd Chern number phases.
This observation emphasizes the role of the SC phases and the vortices induced by the magnetic field.

\bibliography{library}

\begin{thebibliography}{42}%
\makeatletter
\providecommand \@ifxundefined [1]{%
 \@ifx{#1\undefined}
}%
\providecommand \@ifnum [1]{%
 \ifnum #1\expandafter \@firstoftwo
 \else \expandafter \@secondoftwo
 \fi
}%
\providecommand \@ifx [1]{%
 \ifx #1\expandafter \@firstoftwo
 \else \expandafter \@secondoftwo
 \fi
}%
\providecommand \natexlab [1]{#1}%
\providecommand \enquote  [1]{``#1''}%
\providecommand \bibnamefont  [1]{#1}%
\providecommand \bibfnamefont [1]{#1}%
\providecommand \citenamefont [1]{#1}%
\providecommand \href@noop [0]{\@secondoftwo}%
\providecommand \href [0]{\begingroup \@sanitize@url \@href}%
\providecommand \@href[1]{\@@startlink{#1}\@@href}%
\providecommand \@@href[1]{\endgroup#1\@@endlink}%
\providecommand \@sanitize@url [0]{\catcode `\\12\catcode `\$12\catcode
  `\&12\catcode `\#12\catcode `\^12\catcode `\_12\catcode `\%12\relax}%
\providecommand \@@startlink[1]{}%
\providecommand \@@endlink[0]{}%
\providecommand \url  [0]{\begingroup\@sanitize@url \@url }%
\providecommand \@url [1]{\endgroup\@href {#1}{\urlprefix }}%
\providecommand \urlprefix  [0]{URL }%
\providecommand \Eprint [0]{\href }%
\providecommand \doibase [0]{http://dx.doi.org/}%
\providecommand \selectlanguage [0]{\@gobble}%
\providecommand \bibinfo  [0]{\@secondoftwo}%
\providecommand \bibfield  [0]{\@secondoftwo}%
\providecommand \translation [1]{[#1]}%
\providecommand \BibitemOpen [0]{}%
\providecommand \bibitemStop [0]{}%
\providecommand \bibitemNoStop [0]{.\EOS\space}%
\providecommand \EOS [0]{\spacefactor3000\relax}%
\providecommand \BibitemShut  [1]{\csname bibitem#1\endcsname}%
\let\auto@bib@innerbib\@empty
%</preamble>
\bibitem [{\citenamefont {Alicea}(2012)}]{alicea_new_2012}%
  \BibitemOpen
  \bibfield  {author} {\bibinfo {author} {\bibfnamefont {J.}~\bibnamefont
  {Alicea}},\ }\href {\doibase 10.1088/0034-4885/75/7/076501} {\bibfield
  {journal} {\bibinfo  {journal} {Reports on Progress in Physics}\ }\textbf
  {\bibinfo {volume} {75}},\ \bibinfo {pages} {076501} (\bibinfo {year}
  {2012})}\BibitemShut {NoStop}%
\bibitem [{\citenamefont {Bernevig}\ and\ \citenamefont
  {Hughes}(2013)}]{bernevig_topological_2013}%
  \BibitemOpen
  \bibfield  {author} {\bibinfo {author} {\bibfnamefont {B.~A.}\ \bibnamefont
  {Bernevig}}\ and\ \bibinfo {author} {\bibfnamefont {T.~L.}\ \bibnamefont
  {Hughes}},\ }\href@noop {} {\emph {\bibinfo {title} {Topological Insulators
  and Topological Superconductors}}}\ (\bibinfo  {publisher} {{Princeton
  university press}},\ \bibinfo {year} {2013})\BibitemShut {NoStop}%
\bibitem [{\citenamefont {Nayak}\ \emph {et~al.}(2008)\citenamefont {Nayak},
  \citenamefont {Simon}, \citenamefont {Stern}, \citenamefont {Freedman},\ and\
  \citenamefont {Das~Sarma}}]{nayak_non-abelian_2008}%
  \BibitemOpen
  \bibfield  {author} {\bibinfo {author} {\bibfnamefont {C.}~\bibnamefont
  {Nayak}}, \bibinfo {author} {\bibfnamefont {S.~H.}\ \bibnamefont {Simon}},
  \bibinfo {author} {\bibfnamefont {A.}~\bibnamefont {Stern}}, \bibinfo
  {author} {\bibfnamefont {M.}~\bibnamefont {Freedman}}, \ and\ \bibinfo
  {author} {\bibfnamefont {S.}~\bibnamefont {Das~Sarma}},\ }\href {\doibase
  10.1103/RevModPhys.80.1083} {\bibfield  {journal} {\bibinfo  {journal}
  {Reviews of Modern Physics}\ }\textbf {\bibinfo {volume} {80}},\ \bibinfo
  {pages} {1083} (\bibinfo {year} {2008})}\BibitemShut {NoStop}%
\bibitem [{\citenamefont {Sau}\ \emph {et~al.}(2010)\citenamefont {Sau},
  \citenamefont {Lutchyn}, \citenamefont {Tewari},\ and\ \citenamefont
  {Das~Sarma}}]{sau_generic_2010}%
  \BibitemOpen
  \bibfield  {author} {\bibinfo {author} {\bibfnamefont {J.~D.}\ \bibnamefont
  {Sau}}, \bibinfo {author} {\bibfnamefont {R.~M.}\ \bibnamefont {Lutchyn}},
  \bibinfo {author} {\bibfnamefont {S.}~\bibnamefont {Tewari}}, \ and\ \bibinfo
  {author} {\bibfnamefont {S.}~\bibnamefont {Das~Sarma}},\ }\href {\doibase
  10.1103/PhysRevLett.104.040502} {\bibfield  {journal} {\bibinfo  {journal}
  {Physical Review Letters}\ }\textbf {\bibinfo {volume} {104}},\ \bibinfo
  {pages} {040502} (\bibinfo {year} {2010})}\BibitemShut {NoStop}%
\bibitem [{\citenamefont {{Palacio-Morales}}\ \emph {et~al.}(2019)\citenamefont
  {{Palacio-Morales}}, \citenamefont {Mascot}, \citenamefont {Cocklin},
  \citenamefont {Kim}, \citenamefont {Rachel}, \citenamefont {Morr},\ and\
  \citenamefont {Wiesendanger}}]{palacio-morales_atomic-scale_2019}%
  \BibitemOpen
  \bibfield  {author} {\bibinfo {author} {\bibfnamefont {A.}~\bibnamefont
  {{Palacio-Morales}}}, \bibinfo {author} {\bibfnamefont {E.}~\bibnamefont
  {Mascot}}, \bibinfo {author} {\bibfnamefont {S.}~\bibnamefont {Cocklin}},
  \bibinfo {author} {\bibfnamefont {H.}~\bibnamefont {Kim}}, \bibinfo {author}
  {\bibfnamefont {S.}~\bibnamefont {Rachel}}, \bibinfo {author} {\bibfnamefont
  {D.~K.}\ \bibnamefont {Morr}}, \ and\ \bibinfo {author} {\bibfnamefont
  {R.}~\bibnamefont {Wiesendanger}},\ }\href {\doibase 10.1126/sciadv.aav6600}
  {\bibfield  {journal} {\bibinfo  {journal} {Science Advances}\ }\textbf
  {\bibinfo {volume} {5}},\ \bibinfo {pages} {eaav6600} (\bibinfo {year}
  {2019})}\BibitemShut {NoStop}%
\bibitem [{\citenamefont {Lutchyn}\ \emph {et~al.}(2010)\citenamefont
  {Lutchyn}, \citenamefont {Sau},\ and\ \citenamefont
  {Das~Sarma}}]{lutchyn_majorana_2010}%
  \BibitemOpen
  \bibfield  {author} {\bibinfo {author} {\bibfnamefont {R.~M.}\ \bibnamefont
  {Lutchyn}}, \bibinfo {author} {\bibfnamefont {J.~D.}\ \bibnamefont {Sau}}, \
  and\ \bibinfo {author} {\bibfnamefont {S.}~\bibnamefont {Das~Sarma}},\ }\href
  {\doibase 10.1103/PhysRevLett.105.077001} {\bibfield  {journal} {\bibinfo
  {journal} {Physical Review Letters}\ }\textbf {\bibinfo {volume} {105}},\
  \bibinfo {pages} {077001} (\bibinfo {year} {2010})}\BibitemShut {NoStop}%
\bibitem [{\citenamefont {Oreg}\ \emph {et~al.}(2010)\citenamefont {Oreg},
  \citenamefont {Refael},\ and\ \citenamefont {{von
  Oppen}}}]{oreg_helical_2010}%
  \BibitemOpen
  \bibfield  {author} {\bibinfo {author} {\bibfnamefont {Y.}~\bibnamefont
  {Oreg}}, \bibinfo {author} {\bibfnamefont {G.}~\bibnamefont {Refael}}, \ and\
  \bibinfo {author} {\bibfnamefont {F.}~\bibnamefont {{von Oppen}}},\ }\href
  {\doibase 10.1103/PhysRevLett.105.177002} {\bibfield  {journal} {\bibinfo
  {journal} {Physical Review Letters}\ }\textbf {\bibinfo {volume} {105}},\
  \bibinfo {pages} {177002} (\bibinfo {year} {2010})}\BibitemShut {NoStop}%
\bibitem [{\citenamefont {Albuquerque}\ \emph {et~al.}(2008)\citenamefont
  {Albuquerque}, \citenamefont {Katzgraber}, \citenamefont {Troyer},\ and\
  \citenamefont {Blatter}}]{albuquerque_engineering_2008}%
  \BibitemOpen
  \bibfield  {author} {\bibinfo {author} {\bibfnamefont {A.~F.}\ \bibnamefont
  {Albuquerque}}, \bibinfo {author} {\bibfnamefont {H.~G.}\ \bibnamefont
  {Katzgraber}}, \bibinfo {author} {\bibfnamefont {M.}~\bibnamefont {Troyer}},
  \ and\ \bibinfo {author} {\bibfnamefont {G.}~\bibnamefont {Blatter}},\ }\href
  {\doibase 10.1103/PhysRevB.78.014503} {\bibfield  {journal} {\bibinfo
  {journal} {Physical Review B}\ }\textbf {\bibinfo {volume} {78}},\ \bibinfo
  {pages} {014503} (\bibinfo {year} {2008})}\BibitemShut {NoStop}%
\bibitem [{\citenamefont {Alicea}(2010)}]{alicea_majorana_2010}%
  \BibitemOpen
  \bibfield  {author} {\bibinfo {author} {\bibfnamefont {J.}~\bibnamefont
  {Alicea}},\ }\href {\doibase 10.1103/PhysRevB.81.125318} {\bibfield
  {journal} {\bibinfo  {journal} {Physical Review B}\ }\textbf {\bibinfo
  {volume} {81}},\ \bibinfo {pages} {125318} (\bibinfo {year}
  {2010})}\BibitemShut {NoStop}%
\bibitem [{\citenamefont {Fu}\ and\ \citenamefont
  {Kane}(2008)}]{fu_superconducting_2008}%
  \BibitemOpen
  \bibfield  {author} {\bibinfo {author} {\bibfnamefont {L.}~\bibnamefont
  {Fu}}\ and\ \bibinfo {author} {\bibfnamefont {C.~L.}\ \bibnamefont {Kane}},\
  }\href {\doibase 10.1103/PhysRevLett.100.096407} {\bibfield  {journal}
  {\bibinfo  {journal} {Physical Review Letters}\ }\textbf {\bibinfo {volume}
  {100}},\ \bibinfo {pages} {096407} (\bibinfo {year} {2008})}\BibitemShut
  {NoStop}%
\bibitem [{\citenamefont {Qi}\ \emph {et~al.}(2010)\citenamefont {Qi},
  \citenamefont {Hughes},\ and\ \citenamefont {Zhang}}]{qi_chiral_2010}%
  \BibitemOpen
  \bibfield  {author} {\bibinfo {author} {\bibfnamefont {X.-L.}\ \bibnamefont
  {Qi}}, \bibinfo {author} {\bibfnamefont {T.~L.}\ \bibnamefont {Hughes}}, \
  and\ \bibinfo {author} {\bibfnamefont {S.-C.}\ \bibnamefont {Zhang}},\ }\href
  {\doibase 10.1103/PhysRevB.82.184516} {\bibfield  {journal} {\bibinfo
  {journal} {Physical Review B}\ }\textbf {\bibinfo {volume} {82}},\ \bibinfo
  {pages} {184516} (\bibinfo {year} {2010})}\BibitemShut {NoStop}%
\bibitem [{Note1()}]{Note1}%
  \BibitemOpen
  \bibinfo {note} {Full shell systems are an exception; see Ref.~\cite
  {lutchyn_topological_2018}}\BibitemShut {NoStop}%
\bibitem [{\citenamefont {v.~Klitzing}\ \emph {et~al.}(1980)\citenamefont
  {v.~Klitzing}, \citenamefont {Dorda},\ and\ \citenamefont
  {Pepper}}]{klitzing_new_1980}%
  \BibitemOpen
  \bibfield  {author} {\bibinfo {author} {\bibfnamefont {K.}~\bibnamefont
  {v.~Klitzing}}, \bibinfo {author} {\bibfnamefont {G.}~\bibnamefont {Dorda}},
  \ and\ \bibinfo {author} {\bibfnamefont {M.}~\bibnamefont {Pepper}},\ }\href
  {\doibase 10.1103/PhysRevLett.45.494} {\bibfield  {journal} {\bibinfo
  {journal} {Physical Review Letters}\ }\textbf {\bibinfo {volume} {45}},\
  \bibinfo {pages} {494} (\bibinfo {year} {1980})}\BibitemShut {NoStop}%
\bibitem [{\citenamefont {Stern}(2008)}]{stern_anyons_2008}%
  \BibitemOpen
  \bibfield  {author} {\bibinfo {author} {\bibfnamefont {A.}~\bibnamefont
  {Stern}},\ }\href {\doibase 10.1016/j.aop.2007.10.008} {\bibfield  {journal}
  {\bibinfo  {journal} {Annals of Physics}\ }\bibinfo {series} {January
  {{Special Issue}} 2008},\ \textbf {\bibinfo {volume} {323}},\ \bibinfo
  {pages} {204} (\bibinfo {year} {2008})}\BibitemShut {NoStop}%
\bibitem [{\citenamefont {Girvin}(1999)}]{girvin_quantum_1999}%
  \BibitemOpen
  \bibfield  {author} {\bibinfo {author} {\bibfnamefont {S.~M.}\ \bibnamefont
  {Girvin}},\ }in\ \href@noop {} {\emph {\bibinfo {booktitle} {Aspects
  Topologiques de La Physique En Basse Dimension. {{Topological}} Aspects of
  Low Dimensional Systems}}},\ \bibinfo {series and number} {Les {{Houches}} -
  {{Ecole}} d'{{Ete}} de {{Physique Theorique}}},\ \bibinfo {editor} {edited
  by\ \bibinfo {editor} {\bibfnamefont {A.}~\bibnamefont {Comtet}}, \bibinfo
  {editor} {\bibfnamefont {T.}~\bibnamefont {Jolic{\oe}ur}}, \bibinfo {editor}
  {\bibfnamefont {S.}~\bibnamefont {Ouvry}}, \ and\ \bibinfo {editor}
  {\bibfnamefont {F.}~\bibnamefont {David}}}\ (\bibinfo  {publisher} {{Springer
  Berlin Heidelberg}},\ \bibinfo {year} {1999})\ pp.\ \bibinfo {pages}
  {53--175}\BibitemShut {NoStop}%
\bibitem [{\citenamefont {Harper}(1955)}]{harper_single_1955}%
  \BibitemOpen
  \bibfield  {author} {\bibinfo {author} {\bibfnamefont {P.~G.}\ \bibnamefont
  {Harper}},\ }\href {\doibase 10.1088/0370-1298/68/10/304} {\bibfield
  {journal} {\bibinfo  {journal} {Proceedings of the Physical Society. Section
  A}\ }\textbf {\bibinfo {volume} {68}},\ \bibinfo {pages} {874} (\bibinfo
  {year} {1955})}\BibitemShut {NoStop}%
\bibitem [{\citenamefont {Hofstadter}(1976)}]{hofstadter_energy_1976}%
  \BibitemOpen
  \bibfield  {author} {\bibinfo {author} {\bibfnamefont {D.~R.}\ \bibnamefont
  {Hofstadter}},\ }\href {\doibase 10.1103/PhysRevB.14.2239} {\bibfield
  {journal} {\bibinfo  {journal} {Physical Review B}\ }\textbf {\bibinfo
  {volume} {14}},\ \bibinfo {pages} {2239} (\bibinfo {year}
  {1976})}\BibitemShut {NoStop}%
\bibitem [{\citenamefont {B{\o}ttcher}\ \emph {et~al.}(2018)\citenamefont
  {B{\o}ttcher}, \citenamefont {Nichele}, \citenamefont {Kjaergaard},
  \citenamefont {Suominen}, \citenamefont {Shabani}, \citenamefont
  {Palmstr{\o}m},\ and\ \citenamefont
  {Marcus}}]{bottcher_superconducting_2018}%
  \BibitemOpen
  \bibfield  {author} {\bibinfo {author} {\bibfnamefont {C.~G.~L.}\
  \bibnamefont {B{\o}ttcher}}, \bibinfo {author} {\bibfnamefont
  {F.}~\bibnamefont {Nichele}}, \bibinfo {author} {\bibfnamefont
  {M.}~\bibnamefont {Kjaergaard}}, \bibinfo {author} {\bibfnamefont {H.~J.}\
  \bibnamefont {Suominen}}, \bibinfo {author} {\bibfnamefont {J.}~\bibnamefont
  {Shabani}}, \bibinfo {author} {\bibfnamefont {C.~J.}\ \bibnamefont
  {Palmstr{\o}m}}, \ and\ \bibinfo {author} {\bibfnamefont {C.~M.}\
  \bibnamefont {Marcus}},\ }\href {\doibase 10.1038/s41567-018-0259-9}
  {\bibfield  {journal} {\bibinfo  {journal} {Nature Physics}\ }\textbf
  {\bibinfo {volume} {14}},\ \bibinfo {pages} {1138} (\bibinfo {year}
  {2018})}\BibitemShut {NoStop}%
\bibitem [{\citenamefont {Aleiner}\ \emph {et~al.}(2002)\citenamefont
  {Aleiner}, \citenamefont {Brouwer},\ and\ \citenamefont
  {Glazman}}]{aleiner_quantum_2002}%
  \BibitemOpen
  \bibfield  {author} {\bibinfo {author} {\bibfnamefont {I.~L.}\ \bibnamefont
  {Aleiner}}, \bibinfo {author} {\bibfnamefont {P.~W.}\ \bibnamefont
  {Brouwer}}, \ and\ \bibinfo {author} {\bibfnamefont {L.~I.}\ \bibnamefont
  {Glazman}},\ }\href {\doibase 10.1016/S0370-1573(01)00063-1} {\bibfield
  {journal} {\bibinfo  {journal} {Physics Reports}\ }\textbf {\bibinfo {volume}
  {358}},\ \bibinfo {pages} {309} (\bibinfo {year} {2002})}\BibitemShut
  {NoStop}%
\bibitem [{\citenamefont {Jos}(2013)}]{jos_40_2013}%
  \BibitemOpen
  \bibfield  {author} {\bibinfo {author} {\bibfnamefont {J.~V.}\ \bibnamefont
  {Jos}},\ }\href@noop {} {\emph {\bibinfo {title} {40 Years of
  {{Berezinskii}}-{{Kosterlitz}}-{{Thouless}} Theory}}}\ (\bibinfo  {publisher}
  {{World Scientific}},\ \bibinfo {year} {2013})\BibitemShut {NoStop}%
\bibitem [{Note2()}]{Note2}%
  \BibitemOpen
  \bibinfo {note} {They can be eliminated if the magnetic field is applied
  in-plane and strong Dresselhaus spin-orbit coupling (like in (111) InSb
  samples~\cite {alicea_majorana_2010}) is present~\cite
  {dresselhaus_spin-orbit_1955,levine_realizing_2017}}\BibitemShut {NoStop}%
\bibitem [{\citenamefont {Altland}\ and\ \citenamefont
  {Zirnbauer}(1997)}]{altland_nonstandard_1997}%
  \BibitemOpen
  \bibfield  {author} {\bibinfo {author} {\bibfnamefont {A.}~\bibnamefont
  {Altland}}\ and\ \bibinfo {author} {\bibfnamefont {M.~R.}\ \bibnamefont
  {Zirnbauer}},\ }\href {\doibase 10.1103/PhysRevB.55.1142} {\bibfield
  {journal} {\bibinfo  {journal} {Physical Review B}\ }\textbf {\bibinfo
  {volume} {55}},\ \bibinfo {pages} {1142} (\bibinfo {year}
  {1997})}\BibitemShut {NoStop}%
\bibitem [{\citenamefont {Schnyder}\ \emph {et~al.}(2008)\citenamefont
  {Schnyder}, \citenamefont {Ryu}, \citenamefont {Furusaki},\ and\
  \citenamefont {Ludwig}}]{schnyder_classification_2008}%
  \BibitemOpen
  \bibfield  {author} {\bibinfo {author} {\bibfnamefont {A.~P.}\ \bibnamefont
  {Schnyder}}, \bibinfo {author} {\bibfnamefont {S.}~\bibnamefont {Ryu}},
  \bibinfo {author} {\bibfnamefont {A.}~\bibnamefont {Furusaki}}, \ and\
  \bibinfo {author} {\bibfnamefont {A.~W.~W.}\ \bibnamefont {Ludwig}},\ }\href
  {\doibase 10.1103/PhysRevB.78.195125} {\bibfield  {journal} {\bibinfo
  {journal} {Physical Review B}\ }\textbf {\bibinfo {volume} {78}},\ \bibinfo
  {pages} {195125} (\bibinfo {year} {2008})}\BibitemShut {NoStop}%
\bibitem [{\citenamefont {Kitaev}(2009)}]{kitaev_periodic_2009}%
  \BibitemOpen
  \bibfield  {author} {\bibinfo {author} {\bibfnamefont {A.}~\bibnamefont
  {Kitaev}},\ }\href {\doibase 10.1063/1.3149495} {\bibfield  {journal}
  {\bibinfo  {journal} {AIP Conference Proceedings}\ }\textbf {\bibinfo
  {volume} {1134}},\ \bibinfo {pages} {22} (\bibinfo {year}
  {2009})}\BibitemShut {NoStop}%
\bibitem [{\citenamefont {Kitaev}(2001)}]{kitaev_unpaired_2001}%
  \BibitemOpen
  \bibfield  {author} {\bibinfo {author} {\bibfnamefont {A.~Y.}\ \bibnamefont
  {Kitaev}},\ }\href {\doibase 10.1070/1063-7869/44/10S/S29} {\bibfield
  {journal} {\bibinfo  {journal} {Physics-Uspekhi}\ }\textbf {\bibinfo {volume}
  {44}},\ \bibinfo {pages} {131} (\bibinfo {year} {2001})}\BibitemShut
  {NoStop}%
\bibitem [{\citenamefont {Peierls}(1933)}]{peierls_zur_1933}%
  \BibitemOpen
  \bibfield  {author} {\bibinfo {author} {\bibfnamefont {R.}~\bibnamefont
  {Peierls}},\ }\href {\doibase 10.1007/BF01342591} {\bibfield  {journal}
  {\bibinfo  {journal} {Zeitschrift f{\"u}r Physik}\ }\textbf {\bibinfo
  {volume} {80}},\ \bibinfo {pages} {763} (\bibinfo {year} {1933})}\BibitemShut
  {NoStop}%
\bibitem [{\citenamefont {Halsey}(1985)}]{halsey_josephson-junction_1985}%
  \BibitemOpen
  \bibfield  {author} {\bibinfo {author} {\bibfnamefont {T.~C.}\ \bibnamefont
  {Halsey}},\ }\href {\doibase 10.1103/PhysRevB.31.5728} {\bibfield  {journal}
  {\bibinfo  {journal} {Physical Review B}\ }\textbf {\bibinfo {volume} {31}},\
  \bibinfo {pages} {5728} (\bibinfo {year} {1985})}\BibitemShut {NoStop}%
\bibitem [{\citenamefont {Fukui}\ \emph {et~al.}(2005)\citenamefont {Fukui},
  \citenamefont {Hatsugai},\ and\ \citenamefont {Suzuki}}]{fukui_chern_2005}%
  \BibitemOpen
  \bibfield  {author} {\bibinfo {author} {\bibfnamefont {T.}~\bibnamefont
  {Fukui}}, \bibinfo {author} {\bibfnamefont {Y.}~\bibnamefont {Hatsugai}}, \
  and\ \bibinfo {author} {\bibfnamefont {H.}~\bibnamefont {Suzuki}},\ }\href
  {\doibase 10.1143/JPSJ.74.1674} {\bibfield  {journal} {\bibinfo  {journal}
  {Journal of the Physical Society of Japan}\ }\textbf {\bibinfo {volume}
  {74}},\ \bibinfo {pages} {1674} (\bibinfo {year} {2005})}\BibitemShut
  {NoStop}%
\bibitem [{\citenamefont {{Yi-Fu}}\ \emph {et~al.}(2013)\citenamefont
  {{Yi-Fu}}, \citenamefont {{Yun-You}}, \citenamefont {Yan}, \citenamefont
  {Li}, \citenamefont {Rui}, \citenamefont {{Dong-Ning}},\ and\ \citenamefont
  {{Ding-Yu}}}]{yi-fu_coupling-matrix_2013}%
  \BibitemOpen
  \bibfield  {author} {\bibinfo {author} {\bibfnamefont {Z.}~\bibnamefont
  {{Yi-Fu}}}, \bibinfo {author} {\bibfnamefont {Y.}~\bibnamefont {{Yun-You}}},
  \bibinfo {author} {\bibfnamefont {J.}~\bibnamefont {Yan}}, \bibinfo {author}
  {\bibfnamefont {S.}~\bibnamefont {Li}}, \bibinfo {author} {\bibfnamefont
  {S.}~\bibnamefont {Rui}}, \bibinfo {author} {\bibfnamefont {S.}~\bibnamefont
  {{Dong-Ning}}}, \ and\ \bibinfo {author} {\bibfnamefont {X.}~\bibnamefont
  {{Ding-Yu}}},\ }\href {\doibase 10.1088/1674-1056/22/11/117312} {\bibfield
  {journal} {\bibinfo  {journal} {Chinese Physics B}\ }\textbf {\bibinfo
  {volume} {22}},\ \bibinfo {pages} {117312} (\bibinfo {year}
  {2013})}\BibitemShut {NoStop}%
\bibitem [{\citenamefont {Loring}\ and\ \citenamefont
  {Hastings}(2011)}]{loring_disordered_2011}%
  \BibitemOpen
  \bibfield  {author} {\bibinfo {author} {\bibfnamefont {T.~A.}\ \bibnamefont
  {Loring}}\ and\ \bibinfo {author} {\bibfnamefont {M.~B.}\ \bibnamefont
  {Hastings}},\ }\href {\doibase 10.1209/0295-5075/92/67004} {\bibfield
  {journal} {\bibinfo  {journal} {EPL (Europhysics Letters)}\ }\textbf
  {\bibinfo {volume} {92}},\ \bibinfo {pages} {67004} (\bibinfo {year}
  {2011})}\BibitemShut {NoStop}%
\bibitem [{\citenamefont {Thouless}\ \emph {et~al.}(1982)\citenamefont
  {Thouless}, \citenamefont {Kohmoto}, \citenamefont {Nightingale},\ and\
  \citenamefont {{den Nijs}}}]{thouless_quantized_1982}%
  \BibitemOpen
  \bibfield  {author} {\bibinfo {author} {\bibfnamefont {D.~J.}\ \bibnamefont
  {Thouless}}, \bibinfo {author} {\bibfnamefont {M.}~\bibnamefont {Kohmoto}},
  \bibinfo {author} {\bibfnamefont {M.~P.}\ \bibnamefont {Nightingale}}, \ and\
  \bibinfo {author} {\bibfnamefont {M.}~\bibnamefont {{den Nijs}}},\ }\href
  {\doibase 10.1103/PhysRevLett.49.405} {\bibfield  {journal} {\bibinfo
  {journal} {Physical Review Letters}\ }\textbf {\bibinfo {volume} {49}},\
  \bibinfo {pages} {405} (\bibinfo {year} {1982})}\BibitemShut {NoStop}%
\bibitem [{\citenamefont {Fulga}\ \emph {et~al.}(2012)\citenamefont {Fulga},
  \citenamefont {Hassler},\ and\ \citenamefont
  {Akhmerov}}]{fulga_scattering_2012}%
  \BibitemOpen
  \bibfield  {author} {\bibinfo {author} {\bibfnamefont {I.~C.}\ \bibnamefont
  {Fulga}}, \bibinfo {author} {\bibfnamefont {F.}~\bibnamefont {Hassler}}, \
  and\ \bibinfo {author} {\bibfnamefont {A.~R.}\ \bibnamefont {Akhmerov}},\
  }\href {\doibase 10.1103/PhysRevB.85.165409} {\bibfield  {journal} {\bibinfo
  {journal} {Physical Review B}\ }\textbf {\bibinfo {volume} {85}},\ \bibinfo
  {pages} {165409} (\bibinfo {year} {2012})}\BibitemShut {NoStop}%
\bibitem [{\citenamefont {Tinkham}(1996)}]{tinkham_introduction_2004}%
  \BibitemOpen
  \bibfield  {author} {\bibinfo {author} {\bibfnamefont {M.}~\bibnamefont
  {Tinkham}},\ }\href@noop {} {\emph {\bibinfo {title} {Introduction to
  Superconductivity}}},\ \bibinfo {edition} {2nd}\ ed.,\ International Series
  in Pure and Applied Physics\ (\bibinfo  {publisher} {{McGraw-Hill}},\
  \bibinfo {address} {{New York}},\ \bibinfo {year} {1996})\BibitemShut
  {NoStop}%
\bibitem [{\citenamefont {Stern}\ and\ \citenamefont
  {Halperin}(2006)}]{stern_proposed_2006}%
  \BibitemOpen
  \bibfield  {author} {\bibinfo {author} {\bibfnamefont {A.}~\bibnamefont
  {Stern}}\ and\ \bibinfo {author} {\bibfnamefont {B.~I.}\ \bibnamefont
  {Halperin}},\ }\href {\doibase 10.1103/PhysRevLett.96.016802} {\bibfield
  {journal} {\bibinfo  {journal} {Physical Review Letters}\ }\textbf {\bibinfo
  {volume} {96}},\ \bibinfo {pages} {016802} (\bibinfo {year}
  {2006})}\BibitemShut {NoStop}%
\bibitem [{\citenamefont {Bonderson}\ \emph {et~al.}(2006)\citenamefont
  {Bonderson}, \citenamefont {Kitaev},\ and\ \citenamefont
  {Shtengel}}]{bonderson_detecting_2006}%
  \BibitemOpen
  \bibfield  {author} {\bibinfo {author} {\bibfnamefont {P.}~\bibnamefont
  {Bonderson}}, \bibinfo {author} {\bibfnamefont {A.}~\bibnamefont {Kitaev}}, \
  and\ \bibinfo {author} {\bibfnamefont {K.}~\bibnamefont {Shtengel}},\ }\href
  {\doibase 10.1103/PhysRevLett.96.016803} {\bibfield  {journal} {\bibinfo
  {journal} {Physical Review Letters}\ }\textbf {\bibinfo {volume} {96}},\
  \bibinfo {pages} {016803} (\bibinfo {year} {2006})}\BibitemShut {NoStop}%
\bibitem [{\citenamefont {Nilsson}\ \emph {et~al.}(2008)\citenamefont
  {Nilsson}, \citenamefont {Akhmerov},\ and\ \citenamefont
  {Beenakker}}]{nilsson_splitting_2008}%
  \BibitemOpen
  \bibfield  {author} {\bibinfo {author} {\bibfnamefont {J.}~\bibnamefont
  {Nilsson}}, \bibinfo {author} {\bibfnamefont {A.~R.}\ \bibnamefont
  {Akhmerov}}, \ and\ \bibinfo {author} {\bibfnamefont {C.~W.~J.}\ \bibnamefont
  {Beenakker}},\ }\href {\doibase 10.1103/PhysRevLett.101.120403} {\bibfield
  {journal} {\bibinfo  {journal} {Physical Review Letters}\ }\textbf {\bibinfo
  {volume} {101}},\ \bibinfo {pages} {120403} (\bibinfo {year}
  {2008})}\BibitemShut {NoStop}%
\bibitem [{\citenamefont {Banerjee}\ \emph {et~al.}(2018)\citenamefont
  {Banerjee}, \citenamefont {Heiblum}, \citenamefont {Umansky}, \citenamefont
  {Feldman}, \citenamefont {Oreg},\ and\ \citenamefont
  {Stern}}]{banerjee_observation_2018}%
  \BibitemOpen
  \bibfield  {author} {\bibinfo {author} {\bibfnamefont {M.}~\bibnamefont
  {Banerjee}}, \bibinfo {author} {\bibfnamefont {M.}~\bibnamefont {Heiblum}},
  \bibinfo {author} {\bibfnamefont {V.}~\bibnamefont {Umansky}}, \bibinfo
  {author} {\bibfnamefont {D.~E.}\ \bibnamefont {Feldman}}, \bibinfo {author}
  {\bibfnamefont {Y.}~\bibnamefont {Oreg}}, \ and\ \bibinfo {author}
  {\bibfnamefont {A.}~\bibnamefont {Stern}},\ }\href {\doibase
  10.1038/s41586-018-0184-1} {\bibfield  {journal} {\bibinfo  {journal}
  {Nature}\ }\textbf {\bibinfo {volume} {559}},\ \bibinfo {pages} {205}
  (\bibinfo {year} {2018})}\BibitemShut {NoStop}%
\bibitem [{\citenamefont {Karmarkar}(1984)}]{karmarkar_new_1984}%
  \BibitemOpen
  \bibfield  {author} {\bibinfo {author} {\bibfnamefont {N.}~\bibnamefont
  {Karmarkar}},\ }in\ \href {\doibase 10.1145/800057.808695} {\emph {\bibinfo
  {booktitle} {Proceedings of the {{Sixteenth Annual ACM Symposium}} on
  {{Theory}} of {{Computing}}}}},\ \bibinfo {series and number} {{{STOC}} '84}\
  (\bibinfo  {publisher} {{ACM}},\ \bibinfo {address} {{New York, NY, USA}},\
  \bibinfo {year} {1984})\ pp.\ \bibinfo {pages} {302--311}\BibitemShut
  {NoStop}%
\bibitem [{\citenamefont {Forsgren}\ \emph {et~al.}(2002)\citenamefont
  {Forsgren}, \citenamefont {Gill},\ and\ \citenamefont
  {Wright}}]{forsgren_interior_2002}%
  \BibitemOpen
  \bibfield  {author} {\bibinfo {author} {\bibfnamefont {A.}~\bibnamefont
  {Forsgren}}, \bibinfo {author} {\bibfnamefont {P.~E.}\ \bibnamefont {Gill}},
  \ and\ \bibinfo {author} {\bibfnamefont {M.~H.}\ \bibnamefont {Wright}},\
  }\href@noop {} {\bibfield  {journal} {\bibinfo  {journal} {SIAM review}\
  }\textbf {\bibinfo {volume} {44}},\ \bibinfo {pages} {525} (\bibinfo {year}
  {2002})}\BibitemShut {NoStop}%
\bibitem [{\citenamefont {Lutchyn}\ \emph {et~al.}(2018)\citenamefont
  {Lutchyn}, \citenamefont {Winkler}, \citenamefont {{van Heck}}, \citenamefont
  {Karzig}, \citenamefont {Flensberg}, \citenamefont {Glazman},\ and\
  \citenamefont {Nayak}}]{lutchyn_topological_2018}%
  \BibitemOpen
  \bibfield  {author} {\bibinfo {author} {\bibfnamefont {R.~M.}\ \bibnamefont
  {Lutchyn}}, \bibinfo {author} {\bibfnamefont {G.~W.}\ \bibnamefont
  {Winkler}}, \bibinfo {author} {\bibfnamefont {B.}~\bibnamefont {{van Heck}}},
  \bibinfo {author} {\bibfnamefont {T.}~\bibnamefont {Karzig}}, \bibinfo
  {author} {\bibfnamefont {K.}~\bibnamefont {Flensberg}}, \bibinfo {author}
  {\bibfnamefont {L.~I.}\ \bibnamefont {Glazman}}, \ and\ \bibinfo {author}
  {\bibfnamefont {C.}~\bibnamefont {Nayak}},\ }\href@noop {} {\bibfield
  {journal} {\bibinfo  {journal} {arXiv:1809.05512 [cond-mat]}\ } (\bibinfo
  {year} {2018})},\ \Eprint {http://arxiv.org/abs/1809.05512} {arXiv:1809.05512
  [cond-mat]} \BibitemShut {NoStop}%
\bibitem [{\citenamefont {Dresselhaus}(1955)}]{dresselhaus_spin-orbit_1955}%
  \BibitemOpen
  \bibfield  {author} {\bibinfo {author} {\bibfnamefont {G.}~\bibnamefont
  {Dresselhaus}},\ }\href {\doibase 10.1103/PhysRev.100.580} {\bibfield
  {journal} {\bibinfo  {journal} {Physical Review}\ }\textbf {\bibinfo {volume}
  {100}},\ \bibinfo {pages} {580} (\bibinfo {year} {1955})}\BibitemShut
  {NoStop}%
\bibitem [{\citenamefont {Levine}\ \emph {et~al.}(2017)\citenamefont {Levine},
  \citenamefont {Haim},\ and\ \citenamefont {Oreg}}]{levine_realizing_2017}%
  \BibitemOpen
  \bibfield  {author} {\bibinfo {author} {\bibfnamefont {Y.}~\bibnamefont
  {Levine}}, \bibinfo {author} {\bibfnamefont {A.}~\bibnamefont {Haim}}, \ and\
  \bibinfo {author} {\bibfnamefont {Y.}~\bibnamefont {Oreg}},\ }\href {\doibase
  10.1103/PhysRevB.96.165147} {\bibfield  {journal} {\bibinfo  {journal}
  {Physical Review B}\ }\textbf {\bibinfo {volume} {96}},\ \bibinfo {pages}
  {165147} (\bibinfo {year} {2017})}\BibitemShut {NoStop}%
\end{thebibliography}%

\end{document}